\documentclass{article}
\usepackage{amsfonts}

\usepackage{graphicx}
\usepackage{amsmath}

\newtheorem{theorem}{Theorem}
\newtheorem{acknowledgement}[theorem]{Acknowledgement}

\newenvironment{proof}[1][Proof]{\textbf{#1.} }{\ \rule{0.5em}{0.5em}}
\input{tcilatex}

\begin{document}

\title{Quantum Flows as Markovian Limit of Emission, Absorption and Scattering
Interactions}
\author{John Gough \\
%EndAName
Department of Computing \& Mathematics\\
Nottingham-Trent University, Burton Street,\\
Nottingham NG1\ 4BU, United Kingdom.\\
john.gough@ntu.ac.uk}
\date{Commun. Math. Phys. 254, 489-512, 2005}
\maketitle

\begin{abstract}
We consider a Markovian approximation, of weak coupling type, to an open
system perturbation involving emission, absorption and scattering by
reservoir quanta. The result is the general form for a quantum stochastic
flow driven by creation, annihilation and gauge processes. A weak matrix
limit is established for the convergence of the interaction-picture unitary
to a unitary, adapted quantum stochastic process and of the Heisenberg
dynamics to the corresponding quantum stochastic flow: the convergence
strategy is similar to the quantum functional central limits introduced by
Accardi, Frigerio and Lu$^{\left[ 1\right] }$. The principal terms in the
Dyson series expansions are identified and re-summed after the limit to
obtain explicit quantum stochastic differential equations with renormalized
coefficients. An extension of the Pul\'{e} inequalities$^{\left[ 2\right] }$
allows uniform estimates for the Dyson series expansion for both the unitary
operator and the Heisenberg evolution to be obtained.
\end{abstract}

\section{Introduction}

In the interaction picture, the unitary $U_{t}$ arising from a
time-dependent perturbation $V_{t}$, is given by 
\begin{equation}
U_{t}=\mathbf{\vec{T}}\,\exp \left\{ -i\int_{0}^{t}ds\,V_{s}\right\} 
\tag{1.1}  \label{Dyson}
\end{equation}
where $\mathbf{\vec{T}}$\ is Dyson's time-ordering operation. A principal
aim of quantum field theory is then to obtain a normal-ordered version of $%
U_{t}$. When $V_{t}$ involves a sum of monomials of canonical quantum
fields, we may use Feynman rules to expand $U_{t}$: we associate a vertex to
each monomial, with the number of legs corresponding with the degree; we
then construct the class $\frak{F}$\ of Feynman diagrams consisting of such
vertices with certain legs contracted (internal lines) and the remainder
free (external lines); we then specify a rule for writing down an operator $%
L_{G}\left( t\right) $ which, for each $G\in \frak{F}$, will be a
normal-ordered product of the fields associated to the external lines of $G$%
. We then determine a development of the form $U_{t}=\sum_{G\in \frak{F}%
}L_{G}\left( t\right) $. Now, if $G$ can be decomposed as two disconnected
sub-diagrams $G_{1}$ and $G_{2}$, then $L_{G}=\mathbf{\vec{N}\,}%
L_{G_{1}}L_{G_{2}}$ where $\mathbf{\vec{N}}$ is Wick's normal-ordering
operation. This leads to a second presentation of $U_{t}$: 
\begin{equation}
U_{t}=\mathbf{\vec{N}}\,\exp \left\{ \sum_{G\in \frak{F}_{C}}L_{G}\left(
t\right) \right\}  \tag{1.2}  \label{Wick}
\end{equation}
where $\frak{F}_{C}$ is the class of connected Feynman diagrams.

\bigskip

If, in place of quantum fields, we considered quantum white noises, then the
time-ordered presentation corresponds to a Stratonovich form while the
normal-ordered presentation corresponds to an It\={o} form. Our aim is not
to justify this statement, for which there is ample support \cite
{HudsonStreater},\cite{KreeRaczka},\cite{GREP}, but to prove an asymptotic
result which, effectively, is an analogue of the Wong-Zakai theorem for
classical stochastic processes. The interaction that we shall be interested
in is given below as $\left( 1.8\right) $, and is quadratic in the reservoir
creation/annihilation operator fields $a_{t}^{\pm }\left( \lambda \right) $:
the corresponding connected Feynman diagrams will have at most two legs and
therefore will be linear chains. These describe a reservoir quanta created,
subsequently multiply-scattered (i.e., at several times annihilated and
immediately re-created) and finally reabsorbed: external lines may also be
present.

We shall be interested, not in the S-matrix limit $t\rightarrow \infty $,
but in the more subtle van Hove \cite{van Hove}, or weak coupling, limit
where we rescale time as $t/\lambda ^{2}$ with $\lambda $ a coupling
strength parameter appearing in $V_{t}$ and consider the limit $\lambda
\rightarrow 0$ with $t$ fixed. The fields $a_{t}^{\pm }\left( \lambda
\right) $ will converge, in a sense to be spelled out below, to quantum
white noises: more correctly, integrated versions of these fields converge
to the fundamental quantum stochastic processes of Hudson and
Parthasarathy's theory \cite{HP}. The van Hove limit turns out to have
dominant contribution from Feynman diagrams where there is no overlap in the
time ranges of the individual connected subgraphs: these are the so-called 
\textit{type I} terms. All other terms (\textit{type II}) are suppressed. A
similar feature is observed for the limit of the dynamical flow of
observables.

\subsection{The Classical Wong-Zakai Theorem}

Wong and Zakai \cite{Wong Zakai} studied Langevin type equations driven by
differentiable noises $\xi _{t}\left( \lambda \right) $ having correlation $%
\left\langle \xi _{t}\left( \lambda \right) \xi _{s}\left( \lambda \right)
\right\rangle =\frac{1}{\lambda ^{2}}G\left( \frac{t-s}{\lambda ^{2}}\right) 
$ which became delta-correlated only in the limit $\lambda \rightarrow 0$.
They found that the limit dynamics was described by a stochastic
differential equation taking the same form as the pre-limit equations in the
Stratonovich calculus.

Let us specialize to the flow on a symplectic manifold generated by a random
Hamiltonian 
\begin{equation}
\Upsilon _{t}^{\left( \lambda \right) }=H+\sum_{\alpha }F_{\alpha }\xi
_{t}^{\alpha }\left( \lambda \right)  \tag{1.3}
\end{equation}
where $H$ and $F_{\alpha }$ are smooth functions on phase space and $\xi
_{t}^{\alpha }\left( \lambda \right) $ are differentiable stochastic
processes converging to independent white noises. If $x_{t}^{\left( \lambda
\right) }$ is the phase trajectory starting from $x_{0}$ then the evolution
of functions is $J_{t}^{\left( \lambda \right) }\left( f\right) :=f\left(
x_{t}^{\left( \lambda \right) }\right) $. In the limit $\lambda \rightarrow
0 $ we obtain, in accordance with the Wong-Zakai result, the
Stratonovich-Fisk equation 
\begin{equation}
dJ_{t}\left( \cdot \right) =J_{t}\left\{ \cdot ,H\right\} dt+\sum_{\alpha
}J_{t}\left\{ \cdot ,F_{\alpha }\right\} \circ dB_{t}^{\alpha }  \tag{1.4}
\end{equation}
where $B_{t}^{\alpha }$ are independent Wiener processes and the
differential is of Stratonovich type: here we may view the motion as that
governed by the formal Hamiltonian $\Upsilon _{t}=H+\sum_{\alpha }F_{\alpha
}\xi _{t}^{\alpha }$ where $\xi _{t}^{\alpha }$ are white noises. A general
treatment of these problems using the van Hove limit is well-understood \cite
{FrigGori}. These are the stochastic flows that preserve the Poisson bracket
structure \cite{Bismut}. Averaging with respect to the Wiener measure, we
obtain the dynamical semigroup $\mathbb{E}\left[ J_{t}\left( \cdot \right) %
\right] \equiv \exp \left\{ t\frak{L}\left( \cdot \right) \right\} $. From
the It\={o} calculus, the generator will be the hypo-elliptic operator 
\begin{equation}
\frak{L}\left( \cdot \right) =\sum_{\alpha }\left\{ \left\{ \cdot ,F_{\alpha
}\right\} ,F_{\alpha }\right\} +\left\{ \cdot ,H\right\}  \tag{1.5}
\end{equation}
which is already displayed in H\"{o}rmander form.

\subsection{Quantum Markov Approximations}

It was first suggested by Spohn \cite{Spohn} that the weak coupling limit
should be properly considered as a Markovian limit underscored by a
functional central limit. The rigorous determination of irreversible
semigroup evolutions has been given for specific models \cite{DaviesI},\cite
{Pule}. (A detailed account of the derivation of the master equation for a
class of quantum open systems is given in Davies' book \cite{Davies QTOS}.)
The form of the generator of quantum dynamical semigroups was deduced \cite
{GorKossSud},\cite{Lindblad}using the guiding principle that the semi-group
be completely positive. Hudson and Parthasarathy \cite{HP} subsequently
developed a quantum stochastic calculus giving an It\={o} theory of
integration with respect to Bosonic Fock space processes and demonstrated
how to construct dilations of the quantum dynamical semigroups mentioned
above using a Fock space as auxiliary space.

The program now is to begin with a microscopic model for a system-reservoir
interaction and then obtain by some Markovian limit procedure, such as the
weak coupling limit, a quantum stochastic evolution. It was first noted by
von Waldenfels$\ $\cite{vW} that stochastic models successfully describe the
weak coupling limit regime for the Wigner-Weisskopf atom. Later, Accardi,
Frigerio and Lu \cite{AFL} showed how to do this for an interaction of the
type $\Upsilon _{t}^{\left( \lambda \right) }=E_{10}\otimes a_{t}^{+}\left(
\lambda \right) +E_{01}\otimes a_{t}^{-}\left( \lambda \right) $ where $%
E_{10}$ and $E_{01}$ are bounded, mutually adjoint operators on the system
space $\frak{h}_{S}$ and $a_{t}^{\pm }\left( \lambda \right) $ are
creation/annihilation fields having a correlation 
\begin{equation}
\left\langle a_{t}^{+}\left( \lambda \right) a_{s}^{-}\left( \lambda \right)
\right\rangle =\frac{1}{\lambda ^{2}}G\left( \frac{t-s}{\lambda ^{2}}\right)
\tag{1.6}
\end{equation}
where $G\left( .\right) $ is integrable. In the sense of Schwartz
distributions, we have $\lim_{\lambda \rightarrow 0}\,\left\langle
a_{t}^{-}\left( \lambda \right) a_{s}^{+}\left( \lambda \right)
\right\rangle =\gamma \,\delta \left( t-s\right) $ where $\gamma
=\int_{-\infty }^{+\infty }dt\,G\left( t\right) $ is finite. We shall also
take an interest in the constants 
\begin{equation}
\kappa _{+}:=\int_{0}^{\infty }dt\,G\left( t\right) ,\text{ }\kappa
_{-}:=\int_{-\infty }^{0}dt\,G\left( t\right) \text{ and }%
K:=\int_{0}^{\infty }dt\,\left| G\left( t\right) \right| .  \tag{1.7}
\end{equation}
We shall assume that $G\left( -t\right) =G\left( t\right) ^{\ast }$ so that $%
\kappa _{\pm }\equiv \frac{1}{2}\gamma \pm i\sigma $. Already in \cite{AFL},
several important steps were taken: to begin with, there is the anticipation
of the limit algebraic structure by means of a quantum functional central
limit theorem which captures the long time asymptotic behaviour; secondly,
there is the identification of the principal, \textit{type I}, terms in the
Dyson series which survive the Markovian limit (they are the ones arising
from only time-consecutive two-point contractions); finally, there is a
rigorous estimate of the Dyson series expansion employing an argument due to
Pul\'{e} \cite{Pule}.

\subsection{Statement of the Problem}

Our aim is to extend this result in \cite{AFL} to the more general class of
interactions 
\begin{eqnarray}
\Upsilon _{t}^{\left( \lambda \right) } &=&E_{11}\otimes a_{t}^{+}\left(
\lambda \right) a_{t}^{-}\left( \lambda \right) +E_{10}\otimes
a_{t}^{+}\left( \lambda \right) +E_{01}\otimes a_{t}^{-}\left( \lambda
\right) +E_{00}\otimes 1  \notag \\
&=&E_{\alpha \beta }\otimes \left[ a_{t}^{+}\left( \lambda \right) \right]
^{\alpha }\left[ a_{t}^{-}\left( \lambda \right) \right] ^{\beta } 
\TCItag{1.8}
\end{eqnarray}
(\textit{We introduce the summation convention that when the Greek indices }$%
\alpha ,\beta ,\dots $\textit{\ are repeated then we sum each index over the
values }$0$\textit{\ and }$1$\textit{\ - moreover we understand the index }$%
\alpha $\textit{\ in }$\left[ .\right] ^{\alpha }$\textit{\ to represent a
power.}) We require only the conditions that the system operators $E_{\alpha
\beta }$\ are bounded with $K\left\| E_{11}\right\| <1$, where $K$ is the
constant introduced in $\left( 1.7\right) $.

The interaction includes a scattering term, $E_{11}\otimes a_{t}^{+}\left(
\lambda \right) a_{t}^{-}\left( \lambda \right) $, and a constant term. The
terms involving $E_{01}$ and $E_{10}$ describe the emission and absorption
of reservoir quanta and this component has been employed in models of laser
interactions \cite{SpohnLebowitz}. The constant term is of little
consequence as we shall take it to commute with the free Hamiltonian.
However, the scattering term is highly non-trivial: we have to contend with
emission, multiple scatterings and absorption. This means that the number of
terms in the Dyson series expansion of 
\begin{equation}
U_{t}^{\left( \lambda \right) }=\mathbf{\vec{T}}\,\exp \left\{
-i\int_{0}^{t}ds\,\Upsilon _{s}^{\left( \lambda \right) }\right\}  \tag{1.9}
\end{equation}
grows rapidly (in fact, as the Bell numbers of combinatorics \cite{Riodain}%
). However, we are able to prove a uniform estimate of the Dyson series
expansion by a generalization of the Pul\'{e} inequalities, which we give in
section 7. We are then able to re-sum the series to obtain an adapted,
unitary process $U_{t}$ of Hudson-Parthasarathy type (Theorem 8.1). The type
of limit involved is of a weak character and is often referred to as \textit{%
convergence in matrix elements}.

We show that the Heisenberg evolution $J_{t}^{\left( \lambda \right) }\left(
X\right) =U_{t}^{\left( \lambda \right) \dagger }\left( X\otimes
1_{R}\right) U_{t}^{\left( \lambda \right) }$ likewise converges in weak
matrix elements, for fixed bounded observables $X\in \mathcal{B}\left( \frak{%
h}_{S}\right) $, to $J_{t}\left( X\right) =U_{t}^{\lambda \dagger }\left(
X\otimes 1_{R}\right) U_{t}$ (Theorem 10.1).

We are able to obtain the quantum stochastic differential equations
satisfied by $U_{t}$ and by the flow $J_{t}$. In particular, these equations
will involve a gauge differential (due to the scattering) as well as
creation, annihilation and time. In particular, we compute the Lindblad
generator for the flow. We remark that interactions of the type (1.8) were
considered previously in the case where the coefficients $E_{\alpha \beta }$
were commuting operators \cite{C}, and Fermionic operators \cite{Gncma}. In
the former case, a strong resolvent limit was established for the common
spectral resolution, while in the latter, the anti-commutation relations
kill off all but \textit{type I} terms.

\section{Moments and Cumulants}

Let $\Gamma \left( \frak{h}\right) $ be the (Bose) Fock space over the
one-particle Hilbert $\frak{h}$. The Fock vacuum will be denoted by $\Phi $
and the exponential vector map by $\varepsilon :\frak{h}\mapsto \Gamma
\left( \frak{h}\right) $. As usual $\varepsilon \left( 0\right) =\Phi $. We
denote the creation fields as $A^{+}\left( \cdot \right) $, the annihilation
fields as $A^{-}\left( \cdot \right) $ and the differential second
quantization field as $d\Gamma \left( \cdot \right) $, as\ standard. The
Weyl operator with test function $f$ is $W\left( f\right) :=\exp \left[
A^{+}\left( f\right) -A^{-}\left( f\right) \right] $ and we have the Weyl
map $W\left( \cdot \right) $.

As is well-known, the fields $Q\left( \cdot \right) =A^{+}\left( \cdot
\right) +A^{-}\left( \cdot \right) $ are Gaussian random fields when taken
in the Fock vacuum state. More generally, we have \cite{Partha} 
\begin{gather*}
\left\langle \Phi |\,\exp \left\{ it\left( d\Gamma \left( H\right)
+A^{+}\left( Hf\right) +A^{-}\left( Hf\right) +\left\langle
f|Hf\right\rangle \right) \right\} \,\Phi \right\rangle \\
=\exp \int \left( e^{itx}-1\right) d\mu _{H}^{f}\left( dx\right)
\end{gather*}
where $H$ is self-adjoint on $\frak{h}$ with spectral measure $\mu _{H}^{f}$
for vector state $f\in \frak{h}$. This time, we are dealing with Poissonian
fields. We remark that if $\mu _{H}^{f}=\lambda \delta _{1}$,then we obtain
a random variable with Poisson distribution of intensity $\lambda >0$: 
\begin{equation*}
\exp \left\{ \lambda \left( e^{it}-1\right) \right\} =\sum_{n}\sum_{m}\frac{%
\left( it\right) ^{n}}{n!}S\left( n,m\right) \lambda ^{m}.
\end{equation*}
The coefficients $S\left( n,m\right) =\frac{1}{m!}\sum_{l=1}^{m}\left(
-1\right) ^{l+m}l^{n}\binom{m}{l}$ are well-known combinatorial factors:
they are the Stirling number's of the second kind \cite{Riodain}\ and they
count the number of ways of partitioning a set of $n$ items into $m$
non-empty subsets.

The expansion of Poissonian field moments in terms of cumulants, or more
generally the expansion of Green's functions in terms of their connected
Green's functions, can best be described in the language of partitions \cite
{RotaWallstrom}.

A \textit{partition} of the integers $\left\{ 1,\ldots ,n\right\} $ is a
collection of non-empty, disjoint subsets (called parts) whose union is $%
\left\{ 1,\ldots ,n\right\} $. The set of all such partitions will be
denoted as $\frak{P}_{n}$: there will be $S\left( n,m\right) $ partitions of 
$\left\{ 1,\ldots ,n\right\} $ having exactly $m$ parts and $%
B_{n}=\sum_{m}S\left( n,m\right) $ partitions of $\left\{ 1,\ldots
,n\right\} $ in total. $B_{n}$ are called the Bell numbers \cite{Riodain}.

\bigskip

\noindent \textbf{Lemma (2.1)} \textit{Let} $f_{1},g_{1},\dots
,f_{n},g_{n}\in \frak{h}$. \textit{Then} 
\begin{gather}
\sum_{\alpha ,\beta \in \left\{ 0,1\right\} ^{n}}\left\langle \Phi |\,\left[
A^{+}\left( f_{n}\right) \right] ^{\alpha \left( n\right) }\left[
A^{-}\left( g_{n}\right) \right] ^{\beta \left( n\right) }\cdots \left[
A^{+}\left( f_{1}\right) \right] ^{\alpha \left( 1\right) }\left[
A^{-}\left( g_{1}\right) \right] ^{\beta \left( 1\right) }\Phi \right\rangle
\notag \\
=\sum_{\mathcal{A\in }\frak{P}_{n}}\prod_{\left\{ i\left( 1\right) ,\dots
,i\left( k\right) \right\} \in \mathcal{A}}\left\langle g_{i\left( k\right)
}|f_{i\left( k-1\right) }\right\rangle \cdots \left\langle g_{i\left(
3\right) }|f_{i\left( 2\right) }\right\rangle \left\langle g_{i\left(
2\right) }|f_{i\left( 1\right) }\right\rangle  \tag{2.1}
\end{gather}
\textit{where we take the various sets (parts of the partition) }$\left\{
i\left( 1\right) ,\dots ,i\left( k\right) \right\} \in \mathcal{A}$\textit{\
to be ordered so that }$i\left( 1\right) <i\left( 2\right) <\cdots <i\left(
k\right) $\textit{\ and if the set is a singleton it is given the factor of
unity.}

\begin{proof}
If $\alpha \left( i\right) =0,1,$ then we have the absence, respectively
presence, of the creator $A^{+}\left( f_{i}\right) $. Likewise $\beta \left(
i\right) $ gives the absence or presence of the $i$-th annihilator.
Evidently we must have $\alpha \left( n\right) =0=\beta \left( 1\right) $.

Essentially we have a vacuum expectation of a product of $n$ factors $\left[
A^{+}\left( f_{1}\right) \right] ^{\alpha \left( i\right) }$ $\left[
A^{-}\left( g_{1}\right) \right] ^{\beta \left( i\right) }$ and this
ultimately when put to normal order will be a sum of terms each of which is
a product of pair contractions $\left\langle g_{i}|f_{k}\right\rangle $
where $i>k$. For a given term in the sum we write $i\sim k$ if $\left\langle
g_{i}|f_{k}\right\rangle $ appears. An equivalence relation is determined by
a set of contractions as follows: we always have $i\equiv i$ and, more
generally, we have $i\equiv k$ if there exists a sequence $j\left( 1\right)
,\dots j\left( r\right) $ such that either $i\sim j\left( 1\right) \sim
j\left( 2\right) \sim \cdots j\left( r\right) \sim k$ or $k\sim j\left(
1\right) \sim j\left( 2\right) \sim \cdots j\left( r\right) \sim i$. A
partition $\mathcal{A}$ in $\frak{P}_{n}$ is then obtained by looking at the
equivalence classes. (Singletons are just the unpaired labels.) The
correspondence between the terms in the sum and the elements of $\frak{P}%
_{n} $ is one-to-one and the weight given to a particular partition $%
\mathcal{A}\in \frak{P}_{n}$ is just the product of $\left\langle
g_{i}|f_{k}\right\rangle $'s given in (2.1).
\end{proof}

\bigskip

There is a convenient diagrammatic way to understand the formula (2.1). We
first of all associate\ one of four possible vertices with each component $%
\left[ A^{+}\left( f_{j}\right) \right] ^{\alpha \left( j\right) }\left[
A^{-}\left( g_{j}\right) \right] ^{\beta \left( j\right) }$, $j=1,\cdots ,n$%
, they are, for $\left( \alpha _{j},\beta _{j}\right) =\left( 1,1\right)
,\left( 1,0\right) ,\left( 0,1\right) $ and $\left( 0,0\right) $
respectively,

\bigskip

\begin{center}
\begin{tabular}{llll}
%TCIMACRO{
%\TeXButton{scattering vertex}{\setlength{\unitlength}{.1cm}
%\begin{picture}(10,5)
%\label{picdo}
%
%\put(0,0){\dashbox{0.5}(10,0){ }}
%
%\thicklines
%
%\put(5,0){\circle*{1}}
%\put(0,0){\oval(10,10)[tr]}
%\put(10,0){\oval(10,10)[tl]}
%
%\end{picture}
%} }%
%BeginExpansion
\setlength{\unitlength}{.1cm}
\begin{picture}(10,5)
\label{picdo}

\put(0,0){\dashbox{0.5}(10,0){ }}

\thicklines

\put(5,0){\circle*{1}}
\put(0,0){\oval(10,10)[tr]}
\put(10,0){\oval(10,10)[tl]}

\end{picture}
%
%EndExpansion
& 
%TCIMACRO{
%\TeXButton{creation vertex}{\setlength{\unitlength}{.1cm}
%\begin{picture}(10,5)
%\label{pical}
%
%\put(0,0){\dashbox{0.5}(10,0){ }}
%
%\thicklines
%\put(5,0){\circle*{1}}
%\put(0,0){\oval(10,10)[tr]}
%
%\end{picture}
%} }%
%BeginExpansion
\setlength{\unitlength}{.1cm}
\begin{picture}(10,5)
\label{pical}

\put(0,0){\dashbox{0.5}(10,0){ }}

\thicklines
\put(5,0){\circle*{1}}
\put(0,0){\oval(10,10)[tr]}

\end{picture}
%
%EndExpansion
& 
%TCIMACRO{
%\TeXButton{annihilation vertex}{\setlength{\unitlength}{.1cm}
%\begin{picture}(10,5)
%\label{picb}
%
%\put(0,0){\dashbox{0.5}(10,0){ }}
%
%\thicklines
%
%\put(5,0){\circle*{1}}
%\put(10,0){\oval(10,10)[tl]}
%
%\end{picture}
%} }%
%BeginExpansion
\setlength{\unitlength}{.1cm}
\begin{picture}(10,5)
\label{picb}

\put(0,0){\dashbox{0.5}(10,0){ }}

\thicklines

\put(5,0){\circle*{1}}
\put(10,0){\oval(10,10)[tl]}

\end{picture}
%
%EndExpansion
& 
%TCIMACRO{
%\TeXButton{one-vertex}{\setlength{\unitlength}{.1cm}
%\begin{picture}(10,5)
%\label{picd}
%
%\put(0,0){\dashbox{0.5}(10,0){ }}
%
%\thicklines
%
%\put(5,0){\circle*{1}}
%
%
%\end{picture}
%} }%
%BeginExpansion
\setlength{\unitlength}{.1cm}
\begin{picture}(10,5)
\label{picd}

\put(0,0){\dashbox{0.5}(10,0){ }}

\thicklines

\put(5,0){\circle*{1}}

\end{picture}
%
%EndExpansion
\\ 
Scattering & Emission & Absorption & Neutral
\end{tabular}

\bigskip

Figure 1
\end{center}

\noindent We draw the $n$ vertices in a line and proceed to join up the
emission lines to the absorption lines (pair contractions!). A typical
situation is depicted below:

\begin{center}
%TCIMACRO{
%\TeXButton{picture2}{\setlength{\unitlength}{.1cm}
%\begin{picture}(120,30)
%\label{pic2} 
%\thicklines
%
%\put(10,10){\dashbox{0.5}(100,0){ }}
%
%\put(20,10){\circle*{2}}
%\put(25,10){\circle*{2}}
%\put(30,10){\circle*{2}}
%\put(35,10){\circle*{2}}
%\put(40,10){\circle*{2}}
%\put(45,10){\circle*{2}}
%\put(50,10){\circle*{2}}
%\put(55,10){\circle*{2}}
%\put(60,10){\circle*{2}}
%\put(65,10){\circle*{2}}
%\put(70,10){\circle*{2}}
%\put(75,10){\circle*{2}}
%\put(80,10){\circle*{2}}
%\put(85,10){\circle*{2}}
%\put(90,10){\circle*{2}}
%\put(95,10){\circle*{2}}
%\put(100,10){\circle*{2}}
%
%\put(85,10){\oval(20,20)[t]}
%\put(67.5,10){\oval(15,15)[t]}
%\put(50,10){\oval(20,20)[t]}
%\put(35,10){\oval(10,10)[t]}
%
%\put(28,5){$i(5)$}
%\put(38,5){$i(4)$}
%\put(58,5){$i(3)$}
%\put(73,5){$i(2)$}
%\put(93,5){$i(1)$}
%
%
%\thinlines
%\put(35,10){\oval(30,30)[t]}
%\put(60,10){\oval(10,10)[t]}
%\put(95,10){\oval(10,10)[t]}
%\put(85,10){\oval(10,10)[t]}
%\put(52.5,10){\oval(5,5)[t]}
%
%
%\end{picture}
%}}%
%BeginExpansion
\setlength{\unitlength}{.1cm}
\begin{picture}(120,30)
\label{pic2} 
\thicklines

\put(10,10){\dashbox{0.5}(100,0){ }}

\put(20,10){\circle*{2}}
\put(25,10){\circle*{2}}
\put(30,10){\circle*{2}}
\put(35,10){\circle*{2}}
\put(40,10){\circle*{2}}
\put(45,10){\circle*{2}}
\put(50,10){\circle*{2}}
\put(55,10){\circle*{2}}
\put(60,10){\circle*{2}}
\put(65,10){\circle*{2}}
\put(70,10){\circle*{2}}
\put(75,10){\circle*{2}}
\put(80,10){\circle*{2}}
\put(85,10){\circle*{2}}
\put(90,10){\circle*{2}}
\put(95,10){\circle*{2}}
\put(100,10){\circle*{2}}

\put(85,10){\oval(20,20)[t]}
\put(67.5,10){\oval(15,15)[t]}
\put(50,10){\oval(20,20)[t]}
\put(35,10){\oval(10,10)[t]}

\put(28,5){$i(5)$}
\put(38,5){$i(4)$}
\put(58,5){$i(3)$}
\put(73,5){$i(2)$}
\put(93,5){$i(1)$}

\thinlines
\put(35,10){\oval(30,30)[t]}
\put(60,10){\oval(10,10)[t]}
\put(95,10){\oval(10,10)[t]}
\put(85,10){\oval(10,10)[t]}
\put(52.5,10){\oval(5,5)[t]}

\end{picture}
%
%EndExpansion

Figure 2
\end{center}

Evidently we must again join up all creation and annihilation operators into
pairs; we however get creation, multiple scattering and annihilation as the
rule; otherwise we have a single neutral vertex. In the figure, we can think
of a particle being created at vertex $i\left( 1\right) $ then scattered at $%
i\left( 2\right) ,i\left( 3\right) ,i\left( 4\right) $ successively before
being annihilated at $i\left( 5\right) $. (This component has been
highlighted using thick lines.) Now the argument: each such component
corresponds to a unique part, here $\left\{ i\left( 5\right) ,i\left(
4\right) ,i\left( 3\right) ,i\left( 2\right) ,i\left( 1\right) \right\} $,
having two or more elements; singletons may also occur and these are just
the constant term vertices. Therefore every such diagram corresponds
uniquely to a partition of $\left\{ 1,\dots ,n\right\} $.

We remark that $\left( 2.1\right) $ can be considered as a special case of
the expansion $G\left( x_{1},\cdots ,x_{n}\right) =\sum_{\mathcal{A\in }%
\frak{P}_{n}}\prod_{\left\{ i\left( 1\right) ,\dots ,i\left( k\right)
\right\} \in \mathcal{A}}C\left( x_{i\left( 1\right) },\cdots ,x_{i\left(
k\right) }\right) $ of an $n$-particle Green's function $G$ in terms of the
connected Green's functions $C$.

\bigskip

Let us write $\frak{P}$ for the set $\cup _{n}\frak{P}_{n}$ of finite
partitions. With each partition $\mathcal{A\in }\frak{P}_{n}$ we associate a
sequence of occupation numbers $\mathbf{n}=\left( n_{j}\right)
_{j=1}^{\infty }$ where $n_{j}=0,1,2,\dots $ counts the number of $j$-tuples
making up $\mathcal{A}$. In general, we set 
\begin{equation}
E\left( \mathbf{n}\right) :=\sum_{j}jn_{j},\qquad N\left( \mathbf{n}\right)
:=\sum_{j}n_{j}  \tag{2.2}
\end{equation}
so that if $\mathcal{A\in }\frak{P}_{n}$ leads to sequence $\mathbf{n}$,
then $E\left( \mathbf{n}\right) =n$, while $N\left( \mathbf{n}\right) $
counts the number of parts making up the partition. We shall denote by $%
\frak{P}_{\mathbf{n}}$ the set of all partitions having the same occupation
number sequence $\mathbf{n}$.

Given a partition $\mathcal{A\in }\frak{P}_{n}$ we use the convention $%
q\left( j,k,r\right) $ to label the $r-$th element of the $k-$th $j$-tuple.
A simple example of a partition in $\frak{P}_{\mathbf{n}}$ is given by
selecting in order from $\left\{ 1,2,\dots ,E\left( \mathbf{n}\right)
\right\} $ first of all $n_{1}$ singletons, then $n_{2}$ pairs, then $n_{3}$
triples etc. The labelling for this particular partition will be denoted as $%
\bar{q}\left( .,.,.\right) $ and explicitly we have 
\begin{equation}
\bar{q}\left( j,k,r\right) =\sum_{l<j}l\,n_{l}+\left( k-1\right) n_{j}+r. 
\tag{2.3}
\end{equation}

\bigskip

\noindent \textit{Definition \textbf{(2.2)}: We shall denote by} $\frak{S}_{%
\mathbf{n}}^{0}$ \textit{the collection of Pul\'{e} permutations, that is,} $%
\rho \in \frak{S}_{n}$, $E\left( \mathbf{n}\right) =n$, \textit{such that }$%
q=\rho \circ \bar{q}$\textit{\ again describes a partition in} $\frak{P}_{%
\mathbf{n}}$. \textit{Specifically,} $\frak{S}_{\mathbf{n}}^{0}$ \textit{%
consists of all the permutations }$\rho $\textit{\ for which the following
requirements are met:}

\begin{itemize}
\item[i)]  \textit{the order of the individual }$j$\textit{-tuples is
preserved for each }$j$\textit{\ -}

\begin{equation}
\rho \left( \bar{q}\left( j,k,1\right) \right) <\rho \left( \bar{q}\left(
j,k^{\prime },1\right) \right) \text{\qquad }\forall j,1\leq k<k^{\prime
}\leq n_{j};  \tag{2.4}
\end{equation}

\item[ii)]  \textit{creation always precedes annihilation in time for any
contraction pair - } 
\begin{equation}
\rho \left( \bar{q}\left( j,k,1\right) \right) <\rho \left( \bar{q}\left(
j,k,2\right) \right) <\dots <\rho \left( \bar{q}\left( j,k,j\right) \right) 
\text{\qquad }\forall j,1\leq k\leq n_{j}.  \tag{2.5}
\end{equation}
\end{itemize}

In these notations we may rewrite the result of the lemma (2.1) as:

\bigskip

\noindent \textbf{Lemma (2.3)} \textit{Let }$f_{1},g_{1},\dots
,f_{n},g_{n}\in \frak{h}$\textit{. Then}

\begin{gather}
\sum_{\alpha ,\beta \in \left\{ 0,1\right\} ^{n}}\left\langle \Phi |\,\left[
A^{+}\left( f_{n}\right) \right] ^{\alpha \left( n\right) }\left[
A^{-}\left( g_{n}\right) \right] ^{\beta \left( n\right) }\cdots \left[
A^{+}\left( f_{1}\right) \right] ^{\alpha \left( 1\right) }\left[
A^{-}\left( g_{1}\right) \right] ^{\beta \left( 1\right) }\Phi \right\rangle
\notag \\
=\sum_{\mathbf{n}}^{E\left( \mathbf{n}\right) =n}\sum_{\rho \in \frak{S}_{%
\mathbf{n}}^{0}}\prod_{j\geq
2}\prod_{k=1}^{n_{j}}\prod_{r=1}^{j-1}\left\langle g_{\rho \left( \bar{q}%
\left( j,k,r+1\right) \right) }|f_{\rho \left( \bar{q}\left( j,k,r\right)
\right) }\right\rangle  \tag{2.6}
\end{gather}

To better understand this, we return to our diagram conventions. Given an
arbitrary diagram, we wish to construct the Pul\`{e} permutation putting it
to the basic form. For instance, we might have an initial segment of a
diagram looking like the following:

\bigskip

\begin{center}
%TCIMACRO{
%\TeXButton{Typical-Possion}{\setlength{\unitlength}{.05cm}
%\begin{picture}(110,20)
%\label{picga}
%
%
%\put(5,0){\dashbox{0.5}(100,0){ }}
%
%\thicklines
%
%\put(-4,0){\circle*{1}}
%\put(-1,0){\circle*{1}}
%\put(2,0){\circle*{1}}
%
%
%\put(-4,7.5){\circle*{1}}
%\put(-1,7.5){\circle*{1}}
%\put(2,7.5){\circle*{1}}
%
%\put(-4,12.5){\circle*{1}}
%\put(-1,12.5){\circle*{1}}
%\put(2,12.5){\circle*{1}}
%
%\put(10,0){\circle*{2}}
%\put(20,0){\circle*{2}}
%\put(30,0){\circle*{2}}
%\put(40,0){\circle*{2}}
%\put(50,0){\circle*{2}}
%\put(60,0){\circle*{2}}
%\put(70,0){\circle*{2}}
%\put(80,0){\circle*{2}}
%\put(90,0){\circle*{2}}
%\put(100,0){\circle*{2}}
%
%\put(90,0){\oval(20,10)[t]}
%\put(75,0){\oval(30,15)[t]}
%\put(50,0){\oval(20,10)[t]}
%\put(30,0){\oval(20,10)[t]}
%\put(10,0){\oval(40,15)[tr]}
%\put(10,0){\oval(80,25)[tr]}
%
%\end{picture}
%}}%
%BeginExpansion
\setlength{\unitlength}{.05cm}
\begin{picture}(110,20)
\label{picga}

\put(5,0){\dashbox{0.5}(100,0){ }}

\thicklines

\put(-4,0){\circle*{1}}
\put(-1,0){\circle*{1}}
\put(2,0){\circle*{1}}

\put(-4,7.5){\circle*{1}}
\put(-1,7.5){\circle*{1}}
\put(2,7.5){\circle*{1}}

\put(-4,12.5){\circle*{1}}
\put(-1,12.5){\circle*{1}}
\put(2,12.5){\circle*{1}}

\put(10,0){\circle*{2}}
\put(20,0){\circle*{2}}
\put(30,0){\circle*{2}}
\put(40,0){\circle*{2}}
\put(50,0){\circle*{2}}
\put(60,0){\circle*{2}}
\put(70,0){\circle*{2}}
\put(80,0){\circle*{2}}
\put(90,0){\circle*{2}}
\put(100,0){\circle*{2}}

\put(90,0){\oval(20,10)[t]}
\put(75,0){\oval(30,15)[t]}
\put(50,0){\oval(20,10)[t]}
\put(30,0){\oval(20,10)[t]}
\put(10,0){\oval(40,15)[tr]}
\put(10,0){\oval(80,25)[tr]}

\end{picture}
%
%EndExpansion

\bigskip

Figure 3
\end{center}

There will exist a permutation $\sigma $ of the $n$ vertices which will
reorder the vertices so that we have the singletons first, then the pair
contractions, then the triples, etc., so that we obtain a picture of the
following type

\begin{center}
%TCIMACRO{
%\TeXButton{Ordered-Possion}{\setlength{\unitlength}{.05cm}
%\begin{picture}(150,20)
%\label{picga}
%
%
%\put(5,0){\dashbox{0.5}(140,0){ }}
%
%\put(110,10){\vector(1,0){30}}
%\put(140,10){\vector(-1,0){30}}
%\put(112,15){$n_1$ singletons}
%
%
%\put(70,10){\vector(1,0){30}}
%\put(100,10){\vector(-1,0){30}}
%\put(75,15){$n_2$ pairs}
%
%\put(5,10){\vector(1,0){60}}
%\put(20,15){$n_3$ triples}
%
%\thicklines
%
%\put(-4,0){\circle*{1}}
%\put(-1,0){\circle*{1}}
%\put(2,0){\circle*{1}}
%
%
%\put(-4,10){\circle*{1}}
%\put(-1,10){\circle*{1}}
%\put(2,10){\circle*{1}}
%
%
%\put(10,0){\circle*{2}}
%\put(20,0){\circle*{2}}
%\put(30,0){\circle*{2}}
%\put(40,0){\circle*{2}}
%\put(50,0){\circle*{2}}
%\put(60,0){\circle*{2}}
%\put(70,0){\circle*{2}}
%\put(80,0){\circle*{2}}
%\put(90,0){\circle*{2}}
%\put(100,0){\circle*{2}}
%\put(110,0){\circle*{2}}
%\put(120,0){\circle*{2}}
%\put(130,0){\circle*{2}}
%\put(140,0){\circle*{2}}
%
%\put(15,0){\oval(10,10)[t]}
%\put(25,0){\oval(10,10)[t]}
%\put(45,0){\oval(10,10)[t]}
%\put(55,0){\oval(10,10)[t]}
%\put(75,0){\oval(10,10)[t]}
%\put(95,0){\oval(10,10)[t]}
%
%\end{picture}
%}}%
%BeginExpansion
\setlength{\unitlength}{.05cm}
\begin{picture}(150,20)
\label{picga}

\put(5,0){\dashbox{0.5}(140,0){ }}

\put(110,10){\vector(1,0){30}}
\put(140,10){\vector(-1,0){30}}
\put(112,15){$n_1$ singletons}

\put(70,10){\vector(1,0){30}}
\put(100,10){\vector(-1,0){30}}
\put(75,15){$n_2$ pairs}

\put(5,10){\vector(1,0){60}}
\put(20,15){$n_3$ triples}

\thicklines

\put(-4,0){\circle*{1}}
\put(-1,0){\circle*{1}}
\put(2,0){\circle*{1}}

\put(-4,10){\circle*{1}}
\put(-1,10){\circle*{1}}
\put(2,10){\circle*{1}}

\put(10,0){\circle*{2}}
\put(20,0){\circle*{2}}
\put(30,0){\circle*{2}}
\put(40,0){\circle*{2}}
\put(50,0){\circle*{2}}
\put(60,0){\circle*{2}}
\put(70,0){\circle*{2}}
\put(80,0){\circle*{2}}
\put(90,0){\circle*{2}}
\put(100,0){\circle*{2}}
\put(110,0){\circle*{2}}
\put(120,0){\circle*{2}}
\put(130,0){\circle*{2}}
\put(140,0){\circle*{2}}

\put(15,0){\oval(10,10)[t]}
\put(25,0){\oval(10,10)[t]}
\put(45,0){\oval(10,10)[t]}
\put(55,0){\oval(10,10)[t]}
\put(75,0){\oval(10,10)[t]}
\put(95,0){\oval(10,10)[t]}

\end{picture}
%
%EndExpansion

\bigskip

Figure 4
\end{center}

The permutation is again unique if we retain the induced ordering of the
first emission times for each connected block.

\section{A Microscopic Model}

We shall consider a quantum mechanical system $S$ (state space $\frak{h}_{S}$%
) coupled to a Bose quantum field reservoir $R$ over a one-particle space $%
\frak{h}_{R}^{1}$ (state space $\frak{h}_{R}=\Gamma \left( \frak{h}%
_{R}^{1}\right) $). We shall take the reservoir to be in the Fock vacuum
state $\Phi $. The interaction between the system and the reservoir will be
given by the formal Hamiltonian 
\begin{equation}
H^{\left( \lambda \right) }=H_{S}\otimes 1_{R}+1_{S}\otimes d\Gamma \left(
H_{R}^{1}\right) +H_{\mathrm{Int}}^{\left( \lambda \right) }  \tag{3.1}
\end{equation}
where the operators $H_{S}$ and $H_{R}^{1}$ are self-adjoint and bounded
below on $\frak{h}_{S}$ and $\frak{h}_{R}^{1}$, respectively. The
interaction is taken to be 
\begin{equation}
H_{\mathrm{Int}}^{\left( \lambda \right) }=E_{11}\otimes A^{+}\left(
g\right) A^{-}\left( g\right) +\lambda E_{10}\otimes A^{+}\left( g\right)
+\lambda E_{01}\otimes A^{-}\left( g\right) +\lambda ^{2}E_{00}\otimes 1_{R}
\tag{3.2}
\end{equation}
where $E_{\alpha \beta }$ are bounded operators on $\frak{h}_{S}$ with $%
E_{11}$ and $E_{00}$ self-adjoint and $E_{10}=E_{01}^{\dagger }$. The
operators $A^{+}\left( g\right) $ and $A^{-}\left( g\right) $ are the
creation and annihilation operators with test function $g\in \frak{h}%
_{R}^{1} $. (The parameter $\lambda $ is real and will later emerge as a
rescaling parameter in which we hope to obtain a Markovian limit.) We shall
also assume the following harmonic relations 
\begin{eqnarray}
e^{+i\tau H_{S}}\,E_{\alpha \beta }\,e^{-i\tau H_{S}} &=&e^{i\omega \tau
\left( \beta -\alpha \right) }\,E_{\alpha \beta };  \notag \\
e^{+i\tau H_{R}}\,A_{R}^{\pm }\left( g\right) \,e^{-i\tau H_{R}}
&=&A_{R}^{\pm }\left( \theta _{\tau }g\right) .  \TCItag{3.3}
\end{eqnarray}
where $\left( \theta _{\tau }:\tau \in \mathbb{R}\right) $ will be the
one-parameter group of unitaries on $\frak{h}_{R}^{1}$\ with Stone generator 
$H_{R}^{1}$.

We transfer to the interaction picture with the help of the unitary 
\begin{equation}
U\left( \tau ,\lambda \right) =e^{+i\tau \left( H_{S}\otimes
1_{R}+1_{S}\otimes H_{R}\right) }\,e^{-i\tau H^{\left( \lambda \right) }}. 
\tag{3.4}
\end{equation}
In the weak coupling regime, we are interested in the behaviour at long time
scales $\tau =t/\lambda ^{2}$ and from our earlier specifications we see
that $U_{t}^{\left( \lambda \right) }=U\left( t/\lambda ^{2},\lambda \right) 
$ satisfies the interaction picture Schr\"{o}dinger equation 
\begin{equation}
\frac{\partial }{\partial t}U_{t}^{\left( \lambda \right) }=-i\,\Upsilon
_{t}\left( \lambda \right) \,U_{t}^{\left( \lambda \right) }  \tag{3.5}
\end{equation}
with $\Upsilon _{t}\left( \lambda \right) $ as in $\left( 1.8\right) $. Here
we meet the time-dependent rescaled reservoir fields 
\begin{equation}
a_{t}^{\pm }\left( \lambda \right) :=\frac{1}{\lambda }\,e^{\mp i\omega
t/\lambda ^{2}}\,A^{\pm }\left( \theta _{t/\lambda ^{2}}g\right) .  \tag{3.7}
\end{equation}
Specifically we have $\gamma =\int_{-\infty }^{+\infty }d\tau \,\left\langle
g,e^{i\tau \left( H_{R}^{1}-\omega \right) }g\right\rangle =2\pi
\,\left\langle g,\delta \left( H_{R}^{1}-\omega \right) g\right\rangle $ and 
$\kappa _{+}=\left\langle g,\dfrac{1}{i\left( H_{R}^{1}-\omega
-i0^{+}\right) }g\right\rangle =\frac{1}{2}\gamma -i\,$PV$\left\langle g,%
\dfrac{1}{\left( H_{R}^{1}-\omega \right) }g\right\rangle $ where PV denotes
the principle value part.

\section{Quantum Central Limit}

The limit $\lambda \rightarrow 0$ for the above, the two-point function
becomes delta-correlated. However, it is vital to have a mathematical
framework in which to interpret the limit states and observables.

For convenience we set 
\begin{equation}
\theta _{\tau }^{\omega }:=\exp \left\{ i\tau \left( H_{R}^{1}-\omega
\right) \right\} .  \tag{4.1}
\end{equation}
We assume the existence of a non-zero subspace, $\frak{k}$, of $\frak{h}%
_{R}^{1}$ for which 
\begin{equation*}
\int_{-\infty }^{\infty }\left| \left\langle f_{j},\theta _{\upsilon
}^{\omega }f_{k}\right\rangle \right| du<\infty
\end{equation*}
whenever $f_{j},f_{k}\in \frak{k}$. (In reference \cite{AFL}, explicit
examples of dense subspaces, $\frak{k}$, are given and correspond to
``mass-shell'' Hilbert spaces.) The question of completeness can be
addressed immediately: a sesquilinear form on $\frak{k}$ is defined by 
\begin{equation}
\left( f_{j}|f_{k}\right) :=\int_{-\infty }^{\infty }\left\langle
f_{j},\theta _{\upsilon }^{\omega }f_{k}\right\rangle \,du\equiv 2\pi
\,\left\langle f_{j},\delta \left( H_{R}^{1}-\omega \right)
f_{k}\right\rangle  \tag{4.2}
\end{equation}
and we can quotient out the null elements for this form; the completed
Hilbert space will again be denoted by $\frak{k}$ and $\left( .|.\right) $
will be its inner product. The test vector $g$ appearing in the interaction
must belong to $\frak{k}$ so that the constant $\gamma \equiv \left(
g|g\right) $ is finite.

\bigskip

Let $W\left( \cdot \right) $ be the Weyl map from $\frak{h}_{R}^{1}$ as
before. We now fix $f_{j}\in \frak{k}$ and $0\leq S_{j}<T_{j}<\infty $ for
certain indices $j$ and introduce the rescaled operators 
\begin{equation}
A_{\lambda }^{\pm }\left( j\right) :=\frac{1}{\lambda }%
\int_{S_{j}}^{T_{j}}du\,A_{R}^{\pm }\left( \theta _{u/\lambda ^{2}}^{\omega
}f_{j}\right) ,\;W_{\lambda }\left( j\right) :=W\left( \frac{1}{\lambda }%
\int_{S_{j}}^{T_{j}}du\,\theta _{u/\lambda ^{2}}^{\omega }f_{j}\right) . 
\tag{4.3}
\end{equation}
Note that, with respect to our earlier notations (3.7), if $f_{j}=g$ then $%
A_{\lambda }^{\pm }\left( j\right) \equiv \int_{S_{j}}^{T_{j}}du\,a_{u}^{\pm
}\left( \lambda \right) $. The following result is proved as lemma 3.2 in
Accardi, Frigerio and Lu$\ $\cite{AFL}. We write $1_{\left[ S,T\right] }$
for the characteristic function of an interval $\left[ S,T\right] $.

\bigskip

\noindent \textbf{Lemma (4.1) }\textit{For the fields introduced in (4.3)} 
\begin{equation*}
\lim_{\lambda \rightarrow 0}\left[ A_{\lambda }^{-}\left( j\right)
,A_{\lambda }^{+}\left( k\right) \right] =\left( f_{j}|f_{k}\right)
\;\left\langle 1_{\left[ S_{j},T_{j}\right] },1_{\left[ S_{k},T_{k}\right]
}\right\rangle .
\end{equation*}

\bigskip

The right hand side is the inner product $\left\langle f_{j}\otimes 1_{\left[
S_{j},T_{j}\right] },f_{k}\otimes 1_{\left[ S_{k},T_{k}\right]
}\right\rangle $ on the Hilbert space $\frak{k}\otimes L^{2}\left( \mathbb{R}%
^{+}\right) $. This space is isomorphic in a natural way to the $\frak{k}$
-valued square-integrable functions on $\mathbb{R}^{+}$ and we denote this
space as $L^{2}\left( \mathbb{R}^{+},\frak{k}\right) .$

The appropriate noise space for the limit $\lambda \rightarrow 0$ will in
fact be the Bose Fock space $\Gamma \left( L^{2}\left( \mathbb{R}^{+},\frak{k%
}\right) \right) $. Indeed, we have the following fact proved as theorem 3.4
in \cite{AFL}.

\bigskip

\noindent \textbf{Theorem (4.2) }\textit{Let }$\Psi $\textit{\ be the Fock
vacuum for} $\Gamma \left( L^{2}\left( \mathbb{R}^{+},\frak{k}\right)
\right) $ \textit{and let} $W\left( .\right) $ \textit{denote the\ usual
Weyl mapping from} $L^{2}\left( \mathbb{R}^{+},\frak{k}\right) $\textit{\
into the unitaries on }$\Gamma \left( L^{2}\left( \mathbb{R}^{+},\frak{k}%
\right) \right) $. \textit{Then} 
\begin{equation*}
\lim_{\lambda \rightarrow 0}\left\langle \Phi |\,W_{\lambda }\left( 1\right)
\dots W_{\lambda }\left( k\right) \,\Phi \right\rangle =\left\langle \mathbb{%
\Psi }|\,W\left( f_{1}\otimes 1_{\left[ S_{1},T_{1}\right] }\right) \dots
W\left( f_{k}\otimes 1_{\left[ S_{k},T_{k}\right] }\right) \,\mathbb{\Psi }%
\right\rangle
\end{equation*}
\textit{for arbitrary }$k$\textit{\ and} $f_{j}\in \frak{k}$ \textit{and} $%
0\leq S_{j}<T_{j}<\infty $.

\section{The Dyson Series Expansion of $U_{t}^{\left( \protect\lambda
\right) }$}

The formal Dyson series development $U_{t}^{\left( \lambda \right)
}=\sum_{n=0}^{\infty }\left( -i\right) ^{n}D_{n}\left( t,\lambda \right) $
involves the multiple time integrals 
\begin{equation}
D_{n}\left( t,\lambda \right) =\int_{\Delta _{n}\left( t\right) }ds_{n}\dots
ds_{1}\,\Upsilon _{s_{n}}\left( \lambda \right) \dots \Upsilon
_{s_{1}}\left( \lambda \right) .  \tag{5.1}
\end{equation}
For $\sigma \in \frak{S}_{n}$, we introduce the simplex 
\begin{equation}
\Delta _{n}^{\sigma }\left( t\right) :=\left\{ \left( s_{n},\dots
,s_{1}\right) :t>s_{\sigma \left( n\right) }>\cdots >s_{\sigma \left(
1\right) }>0\right\}  \tag{5.2}
\end{equation}
and $\Delta _{n}\left( t\right) $ in $\left( 5.1\right) $ is the simplex
corresponding to the identity permutation.

We consider matrix elements of the type $\left\langle \phi _{1}\otimes
W_{\lambda }\left( 1\right) \Phi |\,U_{t}^{\left( \lambda \right) }\,\phi
_{2}\otimes W_{\lambda }\left( 2\right) \Phi \right\rangle $ with $\phi
_{j}\in \frak{h}_{S}$ and $W_{\lambda }\left( j\right) $ as in $\left(
4.3\right) $. Substituting for the Dyson series, we find that the $n$-th
term can be rewritten as an expectation involving the vacuum state $\Phi $
only:

\begin{eqnarray}
&&\left\langle \phi _{1}\otimes W_{\lambda }\left( 1\right) \Phi |\;\Upsilon
_{s_{n}}\left( \lambda \right) \dots \Upsilon _{s_{1}}\left( \lambda \right)
\;\phi _{2}\otimes W_{\lambda }\left( 2\right) \Phi \right\rangle  \notag \\
&=&\left\langle \phi _{1}\otimes \Phi |\;\tilde{\Upsilon}_{s_{n}}\left(
\lambda \right) \dots \tilde{\Upsilon}_{s_{1}}\left( \lambda \right) \;\phi
_{2}\otimes \Phi \right\rangle \,\left\langle W_{\lambda }\left( 1\right)
\Phi |W_{\lambda }\left( 2\right) \Phi \right\rangle  \TCItag{5.3}
\end{eqnarray}
where $\tilde{\Upsilon}_{s}\left( \lambda \right) $ is obtained from $%
\Upsilon _{s}\left( \lambda \right) $ by the canonical translations 
\begin{equation}
a_{t}^{+}\left( \lambda \right) \rightarrow a_{t}^{+}\left( \lambda \right)
+h_{1}\left( t,\lambda \right) ;\;a_{t}^{-}\left( \lambda \right)
\rightarrow a_{t}^{-}\left( \lambda \right) +h_{2}^{\ast }\left( t,\lambda
\right)  \tag{5.4}
\end{equation}
with

\begin{equation}
h_{j}\left( t,\lambda \right) =\frac{1}{\lambda ^{2}}\int_{S_{j}}^{T_{j}}du%
\,\left\langle \theta _{u/\lambda ^{2}}^{\omega }f_{j}|\theta _{t/\lambda
^{2}}^{\omega }g\right\rangle .  \tag{5.5}
\end{equation}
That is, 
\begin{equation}
\tilde{\Upsilon}_{s}\left( \lambda \right) =\tilde{E}_{\alpha \beta }\left(
t,\lambda \right) \otimes \left[ a_{t}^{+}\left( \lambda \right) \right]
^{\alpha }\left[ a_{t}^{-}\left( \lambda \right) \right] ^{\beta }  \tag{5.6}
\end{equation}
where 
\begin{eqnarray}
\tilde{E}_{00}\left( t,\lambda \right) &=&E_{00}+E_{10}h_{1}\left( t,\lambda
\right) +E_{01}h_{2}^{\ast }\left( t,\lambda \right) +E_{11}h_{1}\left(
t,\lambda \right) h_{2}^{\ast }\left( t,\lambda \right) ;  \notag \\
\tilde{E}_{10}\left( t,\lambda \right) &=&E_{10}+h_{2}^{\ast }\left(
t,\lambda \right) E_{11};  \notag \\
\tilde{E}_{01}\left( t,\lambda \right) &=&E_{01}+h_{1}\left( t,\lambda
\right) E_{11};  \notag \\
\tilde{E}_{11}\left( t,\lambda \right) &=&E_{11}.  \TCItag{5.7}
\end{eqnarray}

In this way we see that the $n$-th term in the Dyson series expansion of the
matrix element is, up to the factor $\left( -i\right) ^{n}\left\langle
W_{\lambda }\left( 1\right) \Phi |W_{\lambda }\left( 2\right) \Phi
\right\rangle $,

\begin{eqnarray}
&&\int_{\Delta _{n}\left( t\right) }ds_{n}\dots ds_{1}\,\left\langle \phi
_{1}|\;\tilde{E}_{\alpha _{n}\beta _{n}}\left( s_{n},\lambda \right) \dots 
\tilde{E}_{\alpha _{1}\beta _{1}}\left( s_{1},\lambda \right) \;\phi
_{2}\right\rangle  \notag \\
&&\times \left\langle \Phi |\,\left[ a_{s_{n}}^{+}\left( \lambda \right) %
\right] ^{\alpha _{n}}\left[ a_{s_{n}}^{-}\left( \lambda \right) \right]
^{\beta _{n}}\dots \left[ a_{s_{1}}^{+}\left( \lambda \right) \right]
^{\alpha _{1}}\left[ a_{s_{1}}^{-}\left( \lambda \right) \right] ^{\beta
_{1}}\,\Phi \right\rangle  \TCItag{5.8}
\end{eqnarray}
and our summation convention is now in place. The vacuum expectation\ can be
computed using lemmas (2.1) or (2.3). The resulting terms can be split into
two types: \textit{type I} will survive the $\lambda \rightarrow 0$ limit; 
\textit{type II} will not. They are distinguished as follows:

\bigskip

\begin{itemize}
\item[\textit{Type I:}]  Terms involving contractions of time consecutive
annihilator-creator pairs only. (That is, under the time-ordered integral in
(5.8), an annihilator $a_{s_{j+1}}^{-}\left( \lambda \right) $ must be
contracted with the creator $a_{s_{j}}^{+}\left( \lambda \right) $.)

\item[\textit{Type II:}]  All others cases.
\end{itemize}

The terminology used here is due to Accardi, Frigerio and Lu \cite{AFL}.

\bigskip

We again resort to a diagrammatic convention in order to describe the Dyson
series expansion into sums of integrals of products of two-point functions.
There is a one-to-one correspondence between the diagrams appearing in the $%
n $-th term of the Dyson series and set of partitions of the $n$ vertices.
The diagram pictured as a typical situation in that section would contribute
a weight of 
\begin{eqnarray*}
&&\left( -i\right) ^{17}\int_{\Delta _{17}\left( t\right) }\tilde{E}%
^{01}\left( t_{17},\lambda \right) \tilde{E}^{00}\left( t_{16},\lambda
\right) \cdots \tilde{E}^{10}\left( t_{1},\lambda \right) \\
&&\times C_{\lambda }\left( t_{17}-t_{11}\right) \cdots C_{\lambda }\left(
t_{2}-t_{1}\right)
\end{eqnarray*}
to the series. Let us consider a typical diagram. We shall assume that
within the diagram there are $n_{1}$ singleton vertices [$\cdots $%
%TCIMACRO{
%\TeXButton{one-vertexa}{\setlength{\unitlength}{.1cm}
%\begin{picture}(10,5)
%\label{picea}
%
%\put(0,0){\dashbox{0.5}(10,0){ }}
%
%\thicklines
%
%\put(5,0){\circle*{1}}
%
%
%\end{picture}
%}}%
%BeginExpansion
\setlength{\unitlength}{.1cm}
\begin{picture}(10,5)
\label{picea}

\put(0,0){\dashbox{0.5}(10,0){ }}

\thicklines

\put(5,0){\circle*{1}}

\end{picture}
%
%EndExpansion
$\cdots ]$, $n_{2}$ contraction pairs [$\cdots $%
%TCIMACRO{
%\TeXButton{contractiona}{\setlength{\unitlength}{.05cm}
%\begin{picture}(26,5)
%\label{picfa}
%
%\put(0,0){\dashbox{0.5}(7,0){ }}
%\put(19,0){\dashbox{0.5}(7,0){ }}
%\put(9,0){$\cdots$}
%\thicklines
%
%\put(5,0){\circle*{2}}
%\put(21,0){\circle*{2}}
%\put(13,0){\oval(16,16)[t]}
%
%\end{picture}
%}}%
%BeginExpansion
\setlength{\unitlength}{.05cm}
\begin{picture}(26,5)
\label{picfa}

\put(0,0){\dashbox{0.5}(7,0){ }}
\put(19,0){\dashbox{0.5}(7,0){ }}
\put(9,0){$\cdots$}
\thicklines

\put(5,0){\circle*{2}}
\put(21,0){\circle*{2}}
\put(13,0){\oval(16,16)[t]}

\end{picture}
%
%EndExpansion
$\cdots ]$, $n_{3}$ contraction triples [$\cdots $%
%TCIMACRO{
%\TeXButton{two-contractiona}{\setlength{\unitlength}{.05cm}
%\begin{picture}(42,8)
%\label{picga}
%
%
%\put(0,0){\dashbox{0.5}(7,0){ }}
%\put(19,0){\dashbox{0.5}(4,0){ }}
%\put(35,0){\dashbox{0.5}(7,0){ }}
%\put(9,0){$\cdots$}
%\put(25,0){$\cdots$}
%\thicklines
%
%\put(5,0){\circle*{2}}
%\put(21,0){\circle*{2}}
%\put(37,0){\circle*{2}}
%
%\put(13,0){\oval(16,16)[t]}
%\put(29,0){\oval(16,16)[t]}
%
%\end{picture}
%}}%
%BeginExpansion
\setlength{\unitlength}{.05cm}
\begin{picture}(42,8)
\label{picga}

\put(0,0){\dashbox{0.5}(7,0){ }}
\put(19,0){\dashbox{0.5}(4,0){ }}
\put(35,0){\dashbox{0.5}(7,0){ }}
\put(9,0){$\cdots$}
\put(25,0){$\cdots$}
\thicklines

\put(5,0){\circle*{2}}
\put(21,0){\circle*{2}}
\put(37,0){\circle*{2}}

\put(13,0){\oval(16,16)[t]}
\put(29,0){\oval(16,16)[t]}

\end{picture}
%
%EndExpansion
$\cdots ]$, etc. That is the diagram has a total of $n=\sum_{j}jn_{j}$
vertices which are partitioned into $m=\sum_{j}n_{j}$ connected subdiagrams.
We see that the total number of diagrams contributing to the $n-$th level of
the Dyson series will be given by the Bell number $B_{n}$.

\section{Principal Terms in the Dyson Series}

A standard technique in perturbative quantum field theory and quantum
statistical mechanics is to develop a series expansion and argue on physical
grounds that certain ``principal terms'' will exceed the other terms in
order of magnitude \cite{AGD}. Often it is possible to re-sum the principal
terms to obtain a useful representation of the dominant behaviour.
Mathematically, the problem comes down to showing that the remaining terms
are negligible in the limiting physical regime being considered.

Let $n$ be a positive integer and $m\in \left\{ 0,...,n-1\right\} $. Let $%
\{\left( p_{j},q_{j}\right) \}_{j=1}^{m}$ be contractions pairs over indices 
$\left\{ 1,\dots ,n\right\} $ such that if $P=\{p_{1},...,p_{m}\}$ and $%
Q=\{q_{1},...,q_{m}\}$ then $P$ and $Q$ are both non-degenerate subsets of
size $m$ and we require that $p_{j}<q_{j}$ for each $j$ and that $Q$ be
ordered so that $q_{1}<...<q_{m}$. We understand that $%
(p_{j},q_{j})_{j=1}^{m}$ is \emph{type I} if $q_{j}=p_{j}+1$ for each $j$
and \emph{type II} otherwise. The following result is an extension of lemma
4.2 in Accardi, Frigerio and Lu \cite{AFL} as now $P\cap Q$ need not be
empty.

\bigskip

\noindent \textbf{Lemma (6.1)} \textit{Let }$(p_{j},q_{j})_{j=1}^{m}$\textit{%
\ be a set of }$m$\textit{\ pairs of contractions over indices }$\left\{
1,\dots ,n\right\} $\textit{\ then} 
\begin{equation}
\left| \int_{\Delta _{n}\left( t\right) }ds_{1}\dots
ds_{n}\,\prod_{j=1}^{m}\left\langle a_{s\left( p_{j}\right) }^{-}\left(
\lambda \right) a_{s\left( q_{j}\right) }^{+}\left( \lambda \right)
\right\rangle \right| \leq \frac{\gamma ^{m}t^{n-m}}{(n-m)!}.  \tag{6.1}
\end{equation}
\textit{Moreover, as }$\lambda \rightarrow 0$, 
\begin{equation}
\int_{\Delta _{n}\left( t\right) }ds_{1}\dots
ds_{n}\,\prod_{j=1}^{m}\left\langle a_{s\left( p_{j}\right) }^{-}\left(
\lambda \right) a_{s\left( q_{j}\right) }^{+}\left( \lambda \right)
\right\rangle \rightarrow \left\{ 
\begin{array}{cc}
\frac{\kappa _{+}^{m}t^{n-m}}{\left( n-m\right) !}, & \text{\textit{type I};}
\\ 
0, & \text{\textit{type II}.}
\end{array}
\right.  \tag{6.2}
\end{equation}

\begin{proof}
Let $q=q_{1}$ and set $t\left( q\right) =\left[ s\left( p\right) -s\left(
q\right) \right] /\lambda ^{2}$ then 
\begin{eqnarray*}
&&\left| \int_{\Delta _{n}\left( t\right) }ds_{1}\dots
ds_{n}\,\prod_{j=1}^{m}\left\langle a_{s\left( p_{j}\right) }^{-}\left(
\lambda \right) a_{s\left( q_{j}\right) }^{+}\left( \lambda \right)
\right\rangle \right| = \\
&&\left| \int_{0}^{t}ds\left( 1\right) \dots \int_{0}^{s\left( q-2\right)
}ds\left( q-1\right) \int_{\left[ s\left( p\right) -s\left( q-1\right) %
\right] /\lambda ^{2}}^{s\left( p\right) /\lambda ^{2}}dt\left( q\right)
\int_{0}^{s\left( p\right) -\lambda ^{2}t\left( q\right) }ds\left(
q+1\right) \dots \right. \\
&&\left. \dots \int_{0}^{s\left( n-1\right) }ds\left( n\right)
\;\left\langle g,\,\theta _{t\left( q\right) }^{\omega }\,g\right\rangle
\,\prod_{j=2}^{m}\left\langle a_{s\left( p_{j}\right) }^{-}\left( \lambda
\right) a_{s\left( q_{j}\right) }^{+}\left( \lambda \right) \right\rangle
\right|
\end{eqnarray*}
However, we have that $s\left( p\right) -\lambda ^{2}t\left( p\right)
<s\left( q-1\right) $ and so we obtain the bound 
\begin{eqnarray*}
&&\left| \int_{0}^{t}ds\left( 1\right) \dots \int_{0}^{s\left( q-2\right)
}ds\left( q-1\right) \int_{-\infty }^{\infty }dt\left( q\right)
\int_{0}^{s\left( p\right) -\lambda ^{2}t\left( q\right) }ds\left(
q+1\right) \dots \right. \\
&&\left. \dots \int_{0}^{s\left( n-1\right) }ds\left( n\right)
\;\left\langle g,\,\theta _{t\left( q\right) }^{\omega }\,g\right\rangle
\,\prod_{j=2}^{m}\left\langle a_{s\left( p_{j}\right) }^{-}\left( \lambda
\right) a_{s\left( q_{j}\right) }^{+}\left( \lambda \right) \right\rangle
\right| .
\end{eqnarray*}
And so, working inductively we obtain (6.1).

Suppose now that the pairs are of \textit{type I}, then $p=q-1$ and so the
lower limit of the $t\left( q\right) $-integral is zero. Consequently, we
encounter the sequence of integrals 
\begin{equation*}
\dots \int_{0}^{s\left( q-2\right) }ds\left( q-1\right) \int_{0}^{s\left(
q-1\right) /\lambda ^{2}}dt\left( q\right) \int_{0}^{s\left( q-1\right)
-\lambda ^{2}t\left( q\right) }ds\left( q+1\right) \dots \left\langle
g,\,\theta _{t\left( q\right) }^{\omega }\,g\right\rangle \dots
\end{equation*}
this occurs for each $q$-variable and so we recognize the limit as stated in
(6.2) for \textit{type I} terms.

If the pairs are of \textit{type II}, on the other hand, then let $j=\min
\left\{ k:p_{k}<q_{k}-1\right\} $; setting $q=q_{k}$, we encounter the
sequence of integrals 
\begin{equation*}
\dots \int_{0}^{s\left( q-2\right) }ds\left( q-1\right) \int_{\left[ s\left(
p\right) -s\left( q-1\right) \right] /\lambda ^{2}}^{s\left( p\right)
/\lambda ^{2}}dt\left( q\right) \int_{0}^{s\left( q-1\right) }ds\left(
q+1\right) \dots \left\langle g,\,\theta _{t\left( q\right) }^{\omega
}\,g\right\rangle \dots
\end{equation*}
but now, with respect to the variables $s(1),...,s(p),...,s(q-1)$ we have
that, since $s\left( p\right) \neq s\left( q-1\right) $, the lower limit $%
\left[ s\left( p\right) -s\left( q-1\right) \right] /\lambda ^{2}$ of the $%
t\left( q\right) $-integral is almost always negative and so, as $t\mapsto
\left\langle g,\,\theta _{t}^{\omega }\,g\right\rangle $ is continuous, we
have the dominated convergence of the whole term to zero.
\end{proof}

\bigskip

Clearly \textit{type II} terms do not contribute to the $n-$th term in the
series expansion in the limit. However, we must establish a uniform bound
for all these terms when the sum over all terms is considered. We do this in
the next section.

Before proceeding let us remark that the expression (5.8) is bounded by 
\begin{gather}
C_{\alpha _{n}\beta _{n}}\dots C_{\alpha _{1}\beta _{1}}\;\left\| \phi
_{1}\right\| \left\| \phi _{2}\right\|  \notag \\
\times \int_{\Delta _{n}\left( t\right) }ds_{n}\dots ds_{1}\left\langle \Phi
|\left[ a_{s_{n}}^{+}\left( \lambda \right) \right] ^{\alpha _{n}}\left[
a_{s_{n}}^{-}\left( \lambda \right) \right] ^{\beta _{n}}\dots \left[
a_{s_{1}}^{+}\left( \lambda \right) \right] ^{\alpha _{1}}\left[
a_{s_{1}}^{-}\left( \lambda \right) \right] ^{\beta _{1}}\Phi \right\rangle 
\notag \\
\tag{6.3}
\end{gather}
where 
\begin{eqnarray}
C_{11} &=&\left\| E_{11}\right\| ;  \notag \\
C_{10} &=&\left\| E_{10}\right\| +\left\| E_{11}\right\|
h_{2};\;C_{01}=\left\| E_{01}\right\| +\left\| E_{11}\right\| h_{1};  \notag
\\
C_{00} &=&\left\| E_{00}\right\| +\left\| E_{10}\right\| h_{1}+\left\|
E_{01}\right\| h_{2}+\left\| E_{11}\right\| h_{1}h_{2}  \TCItag{6.4}
\end{eqnarray}
and $h_{1}=\int_{-\infty }^{\infty }du\,\left| \left\langle g|\theta
_{u}^{\omega }f_{1}\right\rangle \right| $, $h_{2}=\int_{-\infty }^{\infty
}du\,\left| \left\langle g|\theta _{u}^{\omega }f_{2}\right\rangle \right| $.

Recall that we require that $KC_{11}<1$ and\ that $C=\max \left\{
C_{11},C_{10},C_{01},C_{00}\right\} <\infty $.

We need to do some preliminary estimation. We employ the occupation numbers
introduced in section 2. The number of times that we will have $\left(
\alpha ,\beta \right) =\left( 1,1\right) $ in a particular term will be $%
\sum_{j>2}\left( j-2\right) n_{j}$ (that is, singletons and pairs have none,
triples have one, quadruples have two, etc.) and this equals $E\left( 
\mathbf{n}\right) -2N\left( \mathbf{n}\right) +n_{1}$. Therefore, we shall
have 
\begin{equation}
C_{\alpha _{n}\beta _{n}}\dots C_{\alpha _{1}\beta _{1}}\leq C_{11}^{E\left( 
\mathbf{n}\right) -2N\left( \mathbf{n}\right) +n_{1}}C^{2N\left( \mathbf{n}%
\right) -n_{1}}.  \tag{6.5}
\end{equation}

\section{Generalized Pul\'{e} Inequalities}

Putting all this together we get the bound 
\begin{eqnarray}
&&C_{\alpha _{n}\beta _{n}}\dots C_{\alpha _{1}\beta _{1}}\;\int_{\Delta
_{n}\left( t\right) }ds_{n}\dots ds_{1}  \notag \\
&&\times \left\langle \Phi |\,\left[ a_{s_{n}}^{+}\left( \lambda \right) %
\right] ^{\alpha _{n}}\left[ a_{s_{n}}^{-}\left( \lambda \right) \right]
^{\beta _{n}}\dots \left[ a_{s_{1}}^{+}\left( \lambda \right) \right]
^{\alpha _{1}}\left[ a_{s_{1}}^{-}\left( \lambda \right) \right] ^{\beta
_{1}}\,\Phi \right\rangle  \notag \\
&\leq &\sum_{\mathbf{n}}^{E\left( \mathbf{n}\right) =n}\sum_{\rho \in \frak{S%
}_{\mathbf{n}}^{0}}C_{11}^{E\left( \mathbf{n}\right) -2N\left( \mathbf{n}%
\right) +n_{1}}C^{2N\left( \mathbf{n}\right) -n_{1}}\;  \notag \\
&&\times \int_{\Delta _{n}\left( t\right) }ds_{n}\dots ds_{1}\;\prod_{j\geq
2}\prod_{k=1}^{n_{j}}\prod_{r=1}^{j-1}G_{\lambda }\left( s_{\rho \left( \bar{%
q}\left( j,k,r+1\right) \right) }-s_{\rho \left( \bar{q}\left( j,k,r\right)
\right) }\right)  \TCItag{7.1}
\end{eqnarray}
where we use the estimate (6.5) and we obtain the sum over all relevant
terms by summing over all admissible permutations of the basic $\bar{q}$
term. To estimate the simplicial integral we generalize an argument due to
Pul\'{e} (lemma 3 of \cite{Pule}). Let $\tilde{\rho}$ be the induced mapping
on $\mathbb{R}^{n}$ obtained by permuting the Cartesian coordinates
according to $\rho \in \frak{S}_{\mathbf{n}}^{0}$. Then the bound in (7.1)
can be written as 
\begin{eqnarray}
&&\sum_{\mathbf{n}}^{E\left( \mathbf{n}\right) =n}C_{11}^{E\left( \mathbf{n}%
\right) -2N\left( \mathbf{n}\right) +n_{1}}C^{2N\left( \mathbf{n}\right)
-n_{1}}  \notag \\
&&\times \int_{R}ds_{n}\dots ds_{1}\;\prod_{j\geq
2}\prod_{k=1}^{n_{j}}\prod_{r=1}^{j-1}G_{\lambda }\left( s_{\bar{q}\left(
j,k,r+1\right) }-s_{\bar{q}\left( j,k,r\right) }\right)  \TCItag{7.2}
\end{eqnarray}
where $R=\cup \left\{ \tilde{\rho}\Delta _{n}\left( t\right) :\rho \in \frak{%
S}_{\mathbf{n}}^{0}\right\} $. This is down to the fact that the image sets $%
\tilde{\rho}\Delta _{n}\left( t\right) $ will be distinct for different $%
\rho \in \frak{S}_{\mathbf{n}}^{0}$. Now the region, $R$, of integration is
a subset of $\left[ 0,t\right] ^{n}$ for which the variables $s_{\bar{q}%
\left( j,k,1\right) }$ are ordered primarily by the index $j$ and
secondarily by the index $k$. Moreover, each of the variables 
\begin{equation}
u_{\bar{q}\left( j,k,r\right) }:=s_{\bar{q}\left( j,k,r+1\right) }-s_{\bar{q}%
\left( j,k,r\right) }  \tag{7.3}
\end{equation}
are positive, $\left( \forall j;k=1,\dots n_{j};r=1,\dots ,j-1\right) $.
(These properties of $R$ are implicit from the choice of the ordering $\bar{q%
}$ and of the nature of the permutations $\rho \in \frak{S}_{\mathbf{n}}^{0}$%
.) Consider the change of variables 
\begin{equation}
\left( s_{1},\dots ,s_{n}\right) \mapsto \left( s_{\bar{q}\left(
j,k,1\right) };u_{\bar{q}\left( j,k,r\right) }\right)  \tag{7.4}
\end{equation}
where the ordering is first by the $j$, second by the $k,$ and for the $u$'s
finally by the $r=1,\dots ,j-1$. This defines a volume-preserving map which
will take $R$ into $\Delta _{n_{1}}\left( t\right) \times \Delta
_{n_{2}}\left( t\right) \times \cdots \times \lbrack 0,\infty
)^{n_{2}}\times \lbrack 0,\infty )^{2n_{3}}\times \cdots $. From this we are
able to find the upper estimate on (7.2) of the form 
\begin{equation*}
\sum_{\mathbf{n}}^{E\left( \mathbf{n}\right) =n}C_{11}^{E\left( \mathbf{n}%
\right) -2N\left( \mathbf{n}\right) +n_{1}}C^{2N\left( \mathbf{n}\right)
-n_{1}}\;\frac{\left( t\vee 1\right) ^{n_{1}}}{n_{1}!}\frac{\left( t\vee
1\right) ^{n_{2}}}{n_{2}!}\cdots \left[ \int_{0}^{\infty }\left| G_{\lambda
}\left( s\right) \right| ds\right] ^{n_{2}+2n_{2}+\cdots }
\end{equation*}
\begin{eqnarray}
&=&\sum_{\mathbf{n}}^{E\left( \mathbf{n}\right) =n}C_{11}^{E\left( \mathbf{n}%
\right) -2N\left( \mathbf{n}\right) +n_{1}}C^{2N\left( \mathbf{n}\right)
-n_{1}}\;\frac{\left( t\vee 1\right) ^{N\left( \mathbf{n}\right) }}{%
n_{1}!n_{2}!\cdots }K^{E\left( \mathbf{n}\right) -N\left( \mathbf{n}\right) }
\notag \\
&\leq &\sum_{\mathbf{n}}^{E\left( \mathbf{n}\right) =n}\frac{e^{AE\left( 
\mathbf{n}\right) +BN\left( \mathbf{n}\right) }}{n_{1}!n_{2}!\cdots } 
\TCItag{7.5}
\end{eqnarray}
where $A=\ln \left( KC_{11}\right) $ and $B=\ln \left( t\vee 1\right) +\ln
\left( C^{2}\vee 1\right) +\ln \left( C_{11}^{-2}\vee 1\right) +\ln \left(
K^{-1}\vee 1\right) $.

The restriction to those sequences $\mathbf{n}$\ with $E\left( \mathbf{n}%
\right) =n$\ can be lifted and the following estimate for the entire series
obtained 
\begin{equation}
\Omega \left( A,B\right) =\sum_{\mathbf{n}}\frac{e^{AE\left( \mathbf{n}%
\right) +BN\left( \mathbf{n}\right) }}{n_{1}!n_{2}!\cdots }%
=\prod_{k=1}^{\infty }\sum_{n_{k}=0}^{\infty }\frac{e^{\left( kA+B\right)
n_{k}}}{n_{k}!}=\exp \left\{ \frac{e^{A+B}}{1-e^{A}}\right\} .  \tag{7.6}
\end{equation}
The manipulations are familiar from, for example, the calculation of the
grand canonical partition function for the free Bose gas \cite{Pathria}. The
requirement for convergence is that $e^{A}<1$, or equivalently, that $%
KC_{11}<1$.

\section{Limit Transition Amplitudes}

We are now ready to re-sum the Dyson series. First of all, observe that the
functions $h_{j}\left( t,\lambda \right) $ defined in (5.5) will have the
limits 
\begin{equation}
h_{j}\left( t\right) :=\lim_{\lambda \rightarrow 0}h_{j}\left( t,\lambda
\right) =1_{\left[ S_{j},T_{j}\right] }\,\left( f_{j}|g\right) .  \tag{8.1}
\end{equation}
Likewise, we obtain $\tilde{E}_{\alpha \beta }\left( t\right) =\lim_{\lambda
\rightarrow 0}\tilde{E}_{\alpha \beta }\left( t,\lambda \right) $ which will
be just the expressions in (5.7) with the $h_{j}\left( t,\lambda \right) $
replaced by their limits. Explicitly, we have 
\begin{eqnarray}
\tilde{E}_{11}\left( t\right) &=&E_{11},\;\tilde{E}_{01}\left( t\right)
=E_{\alpha 1}\left[ h_{1}\left( t\right) \right] ^{\alpha },  \notag \\
\tilde{E}_{10}\left( t\right) &=&E_{1\beta }\left[ h_{2}^{\ast }\left(
t\right) \right] ^{\beta },\;\tilde{E}_{00}\left( t\right) =\left[
h_{1}\left( t\right) \right] ^{\alpha }E_{\alpha \beta }\left[ h_{2}^{\ast
}\left( t\right) \right] ^{\beta }.  \TCItag{8.2}
\end{eqnarray}

Secondly, only \textit{type I} terms will survive the limit. This means
that, for the $n$-th term in the Dyson series, the only sequences $\alpha
_{1},\beta _{1},\alpha _{2},\beta _{2},\cdots ,\alpha _{n},\beta _{n}$
appearing will be those for which $0=\alpha _{n}=\beta _{1}$ and $\beta
_{l}=\alpha _{l+1}$ for $l=1,\dots ,n-1$.

Thirdly, we encounter the following limit of the two point function: $%
G_{\lambda }\left( t-s\right) $. Let $f$ and $g$ be Schwartz functions then
we will have the limit 
\begin{equation*}
\int_{0}^{T}dt_{2}\int_{0}^{t_{2}}dt_{1}\,G_{\lambda }\left(
t_{2}-t_{1}\right) f\left( t_{2}\right) g\left( t_{1}\right) \rightarrow
\kappa _{+}\int_{0}^{T}ds\,f\left( s\right) g\left( s\right) .
\end{equation*}
Therefore, employing lemma (2.3), we find 
\begin{eqnarray}
&&\lim_{\lambda \rightarrow 0}\left\langle \phi _{1}\otimes W_{\lambda
}\left( 1\right) \Phi |\,U\left( t/\lambda ^{2},\lambda \right) \,\phi
_{2}\otimes W_{\lambda }\left( 2\right) \Phi \right\rangle  \notag \\
&=&\left\langle W\left( f_{1}\otimes 1_{\left[ S_{1},T_{1}\right] }\right)
\Psi |\,W\left( f_{2}\otimes 1_{\left[ S_{2},T_{2}\right] }\right) \Psi
\right\rangle  \notag \\
&&\sum_{n}\left( -i\right) ^{n}\int_{\Delta _{n}\left( t\right)
}ds_{n}\cdots ds_{1}\,\prod_{l=1}^{n-1}\left[ \kappa _{+}\frak{d}_{+}\left(
s_{l+1}-s_{l}\right) \right] ^{\beta _{l}}  \notag \\
&&\times \sum_{\beta \in \left\{ 0,1\right\} ^{n-1}}\left\langle \phi _{1}|\,%
\tilde{E}_{0\beta _{n-1}}\left( s_{n}\right) \cdots \tilde{E}_{\beta
_{2}\beta _{1}}\left( s_{2}\right) \tilde{E}_{\beta _{1}0}\left(
s_{1}\right) \,\phi _{2}\right\rangle  \TCItag{8.3}
\end{eqnarray}
where we use the symbol $\frak{d}_{+}$ for a one-sided delta function: $\int 
\frak{d}_{+}\left( t-s\right) f\left( s\right) ds=f\left( t^{+}\right) $.

We now develop this series. Suppose that we have $\beta _{k+1}=0=\beta _{k}$%
, that is, there are no contractions to the $k-$th term, then we encounter
the factor $\tilde{E}_{00}\left( s\right) =\left[ h_{1}\left( s\right) %
\right] ^{\alpha }E_{\alpha \beta }\left[ h_{2}^{\ast }\left( s\right) %
\right] ^{\beta }$ where $s=s_{k}$. Otherwise, if we have contractions on
the terms associated to \emph{consecutive} variables $s_{k+r},\dots
,s_{k+1},s_{k}$ and we assume that $s_{k+r}$ is not paired to $s_{k+r+1}$%
,nor $s_{k}$ to $s_{k-1}$: then we encounter the factor $\tilde{E}%
_{01}\left( s_{k+r}\right) \tilde{E}_{11}\left( s_{k+r-1}\right) \cdots 
\tilde{E}_{11}\left( s_{k+1}\right) \tilde{E}_{10}\left( s_{k}\right) $ with
the variables $s_{k+r},\dots ,s_{k+1},s_{k}$ all forced equal to a common
value $s$, say. This factor will then be $\left[ h_{1}\left( s_{k}\right) %
\right] ^{\alpha }E_{\alpha 1}\left( E_{11}\right) ^{r-2}E_{1\beta }\left[
h_{2}^{\ast }\left( s_{k}\right) \right] ^{\beta }$.

Now (8.3) involves a sum over all consecutive pairings: the corresponding
partition will have all parts consisting of consecutive labels. We can list
these parts in increasing order, say from 1 to $m$ if there are $m$ of them,
and let $r_{j}$ be the size of the $j$-th part. The number of contractions
will be $\sum \beta _{l}$ and this will be $n-m=\sum_{j=1}^{m}\left(
r_{j}-1\right) $. With these observations we see that (8.3) becomes 
\begin{eqnarray}
&&\left\langle W\left( f_{1}\otimes 1_{\left[ S_{1},T_{1}\right] }\right)
\Psi |\,W\left( f_{2}\otimes 1_{\left[ S_{2},T_{2}\right] }\right) \Psi
\right\rangle \;\sum_{n}\sum_{m}\sum_{r_{m},\dots r_{1}\geq 1}^{r_{1}+\cdots
+r_{m}=n}  \notag \\
&&\times \int_{\Delta _{m}\left( t\right) }ds_{m}\cdots ds_{1}\;\left(
-i\right) ^{\sum_{j=1}^{m}r_{j}}\kappa _{+}^{\sum_{j=1}^{m}\left(
r_{j}-1\right) }\;\left\langle \phi _{1}|\,E_{\alpha _{m},\beta
_{m}}^{\left( r_{m}\right) }\cdots E_{\alpha _{1},\beta _{1}}^{\left(
r_{1}\right) }\,\phi _{2}\right\rangle  \notag \\
&&\left[ h_{1}\left( s_{m}\right) \right] ^{\alpha _{m}}\left[ h_{2}^{\ast
}\left( s_{m}\right) \right] ^{\beta _{m}}\cdots \left[ h_{1}\left(
s_{1}\right) \right] ^{\alpha _{1}}\left[ h_{2}^{\ast }\left( s_{1}\right) %
\right] ^{\beta _{1}}  \TCItag{8.4}
\end{eqnarray}
where we set 
\begin{equation}
E_{\alpha ,\beta }^{\left( r\right) }:=\left\{ 
\begin{array}{ll}
E_{\alpha \beta }, & r=1; \\ 
E_{\alpha 1}\left( E_{11}\right) ^{r-2}E_{1\beta }, & r\geq 2.
\end{array}
\right.  \tag{8.5}
\end{equation}
In the following, we shall encounter the coefficients 
\begin{equation}
L_{\alpha \beta }:=-i\sum_{r=1}^{\infty }\left( -i\kappa \right)
^{r-1}E_{\alpha ,\beta }^{\left( r\right) }=-iE_{\alpha \beta }-\kappa
E_{\alpha 1}\frac{1}{1+i\kappa E_{11}}E_{1\beta }.  \tag{8.6}
\end{equation}

\bigskip

With respect to the representation $L^{2}\left( \mathbb{R}^{+},\frak{k}%
\right) \cong \frak{k}\otimes L^{2}\left( \mathbb{R}^{+}\right) $, we
introduce the four fundamental operator processes (here $\chi _{\lbrack
0,t]} $ is the operator on $L^{2}\left( \mathbb{R}^{+}\right) $
corresponding to multiplication by $1_{\left[ 0,t\right] }$) 
\begin{equation}
\begin{tabular}{ll}
(creation) & $A_{t}^{10}=A^{+}\left( g\otimes 1_{\left[ 0,t\right] }\right)
; $ \\ 
(conservation) & $A_{t}^{11}=d\Gamma \left( |g)(g|\otimes \chi _{\lbrack
0,t]}\right) ;$ \\ 
(annihilation) & $A_{t}^{01}=A^{-}\left( g\otimes 1_{\left[ 0,t\right]
}\right) ;$ \\ 
(time) & $A_{t}^{00}=t.$%
\end{tabular}
\tag{8.8}
\end{equation}
These are the basic quantum stochastic processes on the Hudson-Parthasarathy
space $\Gamma \left( L^{2}\left( \mathbb{R}^{+},\frak{k}\right) \right) $.
We note that the quantum It\={o} table takes the concise form 
\begin{equation}
dA_{t}^{\alpha 1}dA_{t}^{1\beta }=\gamma \,dA_{t}^{\alpha \beta }  \tag{8.9}
\end{equation}
with all other pairs vanishing.

\bigskip

\noindent \textbf{Theorem (8.1)} \textit{Suppose the system operators }$%
E_{\alpha \beta }$\textit{\ are bounded with }$K\left\| E_{11}\right\| <1$%
\textit{. Let }$\phi _{1},\phi _{2}\in \frak{h}_{S}$ \textit{and} $%
f_{1},f_{2}\in \frak{k}$. \textit{Then} 
\begin{eqnarray*}
&&\lim_{\lambda \rightarrow 0}\left\langle \phi _{1}\otimes W_{\lambda
}\left( 1\right) \Phi |\,U_{t}^{\left( \lambda \right) }\,\phi _{2}\otimes
W_{\lambda }\left( 2\right) \Phi \right\rangle \\
&=&\left\langle \phi _{1}\otimes W\left( f_{1}\otimes 1_{\left[ S_{1},T_{1}%
\right] }\right) \Psi |\,U_{t}\,\phi _{2}\otimes W\left( f_{2}\otimes 1_{%
\left[ S_{2},T_{2}\right] }\right) \Psi \right\rangle
\end{eqnarray*}
\textit{where }$\left( U_{t}:t\geq 0\right) $\textit{\ is a unitary adapted
quantum stochastic process on }$\frak{h}_{S}\otimes \Gamma \left(
L^{2}\left( \mathbb{R}^{+},\frak{k}\right) \right) $ \textit{satisfying the
quantum stochastic differential equation} 
\begin{equation}
dU_{t}=L_{\alpha \beta }U_{t}\otimes dA_{t}^{\alpha \beta }  \tag{8.10}
\end{equation}
\textit{with }$U_{0}=1$\textit{\ and where the coefficients are given by
(8.6):} 
\begin{eqnarray*}
L_{11} &=&-iE_{11}(1+i\kappa E_{11})^{-1},\quad L_{10}=-i(1+i\kappa
E_{11})^{-1}E_{10} \\
L_{01} &=&-iE_{01}(1+i\kappa E_{11})^{-1},\quad L_{00}=-iE_{00}-\kappa
E_{01}(1+i\kappa E_{11})^{-1}E_{10}.
\end{eqnarray*}

\begin{proof}
The quantum stochastic differential equation (8.10) takes the form 
\begin{eqnarray*}
dU_{t} &=&\frac{1}{\gamma }\left( W-1\right) U_{t}\otimes
dA_{t}^{11}+LU_{t}\otimes dA_{t}^{10} \\
&&-L^{\dagger }WU_{t}\otimes dA_{t}^{01}-\left( \frac{1}{2}\gamma L^{\dagger
}L+iH\right) U_{t}\otimes dA_{t}^{00}
\end{eqnarray*}
where 
\begin{eqnarray}
W &=&\frac{1-i\kappa _{-}E_{11}}{1+i\kappa _{+}E_{11}}\;\text{(unitary)} 
\notag \\
L &=&-i(1+i\kappa _{+}E_{11})^{-1}E_{10}\;\text{(bounded)}  \notag \\
H &=&E_{00}+\func{Im}\left\{ \kappa _{+}E_{01}\frac{1}{1+i\kappa _{+}E_{11}}%
E_{10}\right\} \;\text{(self-adjoint).}  \TCItag{8.11}
\end{eqnarray}
A fundamental result of quantum stochastic calculus \cite{HP} is that the
process $U_{t}$ defined as the solution of $\left( 8.11\right) $ with
initial condition $U_{0}=1$, exists and is an adapted, unitary process. With
our summation convention in place, we have the chaotic expansion 
\begin{equation}
U_{t}=\sum_{m\geq 0}\int_{\Delta _{m}\left( t\right) }L_{\alpha \left(
m\right) \beta \left( m\right) }\cdots L_{\alpha \left( 1\right) \beta
\left( 1\right) }\otimes dA_{s\left( m\right) }^{\alpha \left( m\right)
\beta \left( m\right) }\cdots dA_{s\left( 1\right) }^{\alpha \left( 1\right)
\beta \left( 1\right) }  \tag{8.12}
\end{equation}
and so $\left\langle \phi _{1}\otimes W\left( f_{1}\otimes 1_{\left[
S_{1},T_{1}\right] }\right) \Psi |\,U_{t}\,\phi _{2}\otimes W\left(
f_{2}\otimes 1_{\left[ S_{2},T_{2}\right] }\right) \Psi \right\rangle $ can
be expressed as 
\begin{gather*}
\left\langle W\left( f_{1}\otimes 1_{\left[ S_{1},T_{1}\right] }\right) \Psi
|W\left( f_{2}\otimes 1_{\left[ S_{2},T_{2}\right] }\right) \Psi
\right\rangle \sum_{m\geq 0}\left\langle \phi _{1}|L_{\alpha \left( m\right)
\beta \left( m\right) }\cdots L_{\alpha \left( 2\right) \beta \left(
2\right) }L_{\alpha \left( 1\right) \beta \left( 1\right) }\phi
_{2}\right\rangle \\
\times \int_{\Delta _{m}\left( t\right) }ds_{m}\cdots ds_{1}\,\left( \left[
h_{1}\left( s_{m}\right) \right] ^{\alpha \left( m\right) }\left[
h_{2}^{\ast }\left( s_{m}\right) \right] ^{\beta \left( m\right) }\right)
\cdots \left( \left[ h_{1}\left( s_{1}\right) \right] ^{\alpha \left(
1\right) }\left[ h_{2}^{\ast }\left( s_{1}\right) \right] ^{\beta \left(
1\right) }\right) .
\end{gather*}
By inspection, this evidently agrees with (8.4).
\end{proof}

\subsection{Re-summing the Series}

Again we drop all diagrams that are type II to get the series

\bigskip

\begin{eqnarray*}
%TCIMACRO{
%\TeXButton{vacuum}{\setlength{\unitlength}{.05cm}
%\begin{picture}(20,5)
%\label{picc}
%
%\put(0,0){\dashbox{0.5}(5,0){ }}
%\put(15,0){\dashbox{0.5}(5,0){ }}
%
%\thicklines
%
%\put(10,0){\circle{10}}
%
%
%\end{picture}
%} }%
%BeginExpansion
\setlength{\unitlength}{.05cm}
\begin{picture}(20,5)
\label{picc}

\put(0,0){\dashbox{0.5}(5,0){ }}
\put(15,0){\dashbox{0.5}(5,0){ }}

\thicklines

\put(10,0){\circle{10}}

\end{picture}
%
%EndExpansion
&=&%
%TCIMACRO{
%\TeXButton{identity}{\setlength{\unitlength}{.05cm}
%\begin{picture}(20,5)
%\label{picc}
%
%\put(0,0){\dashbox{0.5}(20,0){ }}
%
%
%\end{picture}
%}}%
%BeginExpansion
\setlength{\unitlength}{.05cm}
\begin{picture}(20,5)
\label{picc}

\put(0,0){\dashbox{0.5}(20,0){ }}

\end{picture}
%
%EndExpansion
+\left[ 
%TCIMACRO{
%\TeXButton{1 neutral}{\setlength{\unitlength}{.05cm}
%\begin{picture}(20,5)
%\label{picc}
%
%\put(0,0){\dashbox{0.5}(20,0){ }}
%
%\thicklines
%\put(10,0){\circle*{2}}
%
%\end{picture}
%}}%
%BeginExpansion
\setlength{\unitlength}{.05cm}
\begin{picture}(20,5)
\label{picc}

\put(0,0){\dashbox{0.5}(20,0){ }}

\thicklines
\put(10,0){\circle*{2}}

\end{picture}
%
%EndExpansion
\right] +\left[ 
%TCIMACRO{
%\TeXButton{2 neutral}{\setlength{\unitlength}{.05cm}
%\begin{picture}(20,5)
%\label{picc}
%
%\put(0,0){\dashbox{0.5}(20,0){ }}
%
%\thicklines
%\put(5,0){\circle*{2}}
%\put(15,0){\circle*{2}}
%
%\end{picture}
%}}%
%BeginExpansion
\setlength{\unitlength}{.05cm}
\begin{picture}(20,5)
\label{picc}

\put(0,0){\dashbox{0.5}(20,0){ }}

\thicklines
\put(5,0){\circle*{2}}
\put(15,0){\circle*{2}}

\end{picture}
%
%EndExpansion
+%
%TCIMACRO{
%\TeXButton{contraction}{\setlength{\unitlength}{.05cm}
%\begin{picture}(20,5)
%\label{picc}
%
%\put(0,0){\dashbox{0.5}(20,0){ }}
%
%\thicklines
%\put(5,0){\circle*{2}}
%\put(15,0){\circle*{2}}
%
%\put(10,0){\oval(10,10)[t]}
%
%\end{picture}
%}}%
%BeginExpansion
\setlength{\unitlength}{.05cm}
\begin{picture}(20,5)
\label{picc}

\put(0,0){\dashbox{0.5}(20,0){ }}

\thicklines
\put(5,0){\circle*{2}}
\put(15,0){\circle*{2}}

\put(10,0){\oval(10,10)[t]}

\end{picture}
%
%EndExpansion
\right] \\
&& \\
&&+\left[ 
%TCIMACRO{
%\TeXButton{2 neutral}{\setlength{\unitlength}{.05cm}
%\begin{picture}(30,5)
%\label{picc}
%
%\put(0,0){\dashbox{0.5}(30,0){ }}
%
%\thicklines
%\put(5,0){\circle*{2}}
%\put(15,0){\circle*{2}}
%\put(25,0){\circle*{2}}
%
%\end{picture}
%}}%
%BeginExpansion
\setlength{\unitlength}{.05cm}
\begin{picture}(30,5)
\label{picc}

\put(0,0){\dashbox{0.5}(30,0){ }}

\thicklines
\put(5,0){\circle*{2}}
\put(15,0){\circle*{2}}
\put(25,0){\circle*{2}}

\end{picture}
%
%EndExpansion
+%
%TCIMACRO{
%\TeXButton{2 neutral}{\setlength{\unitlength}{.05cm}
%\begin{picture}(30,5)
%\label{picc}
%
%\put(0,0){\dashbox{0.5}(30,0){ }}
%
%\thicklines
%\put(5,0){\circle*{2}}
%\put(15,0){\circle*{2}}
%\put(25,0){\circle*{2}}
%
%
%\put(20,0){\oval(10,10)[t]}
%
%\end{picture}
%}}%
%BeginExpansion
\setlength{\unitlength}{.05cm}
\begin{picture}(30,5)
\label{picc}

\put(0,0){\dashbox{0.5}(30,0){ }}

\thicklines
\put(5,0){\circle*{2}}
\put(15,0){\circle*{2}}
\put(25,0){\circle*{2}}

\put(20,0){\oval(10,10)[t]}

\end{picture}
%
%EndExpansion
+%
%TCIMACRO{
%\TeXButton{2 neutral}{\setlength{\unitlength}{.05cm}
%\begin{picture}(30,5)
%\label{picc}
%
%\put(0,0){\dashbox{0.5}(30,0){ }}
%
%\thicklines
%\put(5,0){\circle*{2}}
%\put(15,0){\circle*{2}}
%\put(25,0){\circle*{2}}
%
%\put(10,0){\oval(10,10)[t]}
%
%\end{picture}
%}}%
%BeginExpansion
\setlength{\unitlength}{.05cm}
\begin{picture}(30,5)
\label{picc}

\put(0,0){\dashbox{0.5}(30,0){ }}

\thicklines
\put(5,0){\circle*{2}}
\put(15,0){\circle*{2}}
\put(25,0){\circle*{2}}

\put(10,0){\oval(10,10)[t]}

\end{picture}
%
%EndExpansion
+%
%TCIMACRO{
%\TeXButton{2 neutral}{\setlength{\unitlength}{.05cm}
%\begin{picture}(30,5)
%\label{picc}
%
%\put(0,0){\dashbox{0.5}(30,0){ }}
%
%\thicklines
%\put(5,0){\circle*{2}}
%\put(15,0){\circle*{2}}
%\put(25,0){\circle*{2}}
%
%\put(10,0){\oval(10,10)[t]}
%\put(20,0){\oval(10,10)[t]}
%
%\end{picture}
%}}%
%BeginExpansion
\setlength{\unitlength}{.05cm}
\begin{picture}(30,5)
\label{picc}

\put(0,0){\dashbox{0.5}(30,0){ }}

\thicklines
\put(5,0){\circle*{2}}
\put(15,0){\circle*{2}}
\put(25,0){\circle*{2}}

\put(10,0){\oval(10,10)[t]}
\put(20,0){\oval(10,10)[t]}

\end{picture}
%
%EndExpansion
\right] \\
&& \\
&&+\ \cdots
\end{eqnarray*}

We see the first appearance of scattering in the last term in the 3rd term
of the series: such terms however eventually out-proliferate diagrams with
no scattering. The terms have been grouped by vertex number, however, it
also possible to group them by effective vertex number (equal to the number
of parts, or equivalently the original simplex degree minus the number of
contractions) to give

\begin{center}
%TCIMACRO{
%\TeXButton{vacuum}{\setlength{\unitlength}{.05cm}
%\begin{picture}(20,5)
%\label{picc}
%
%\put(0,0){\dashbox{0.5}(5,0){ }}
%\put(15,0){\dashbox{0.5}(5,0){ }}
%
%\thicklines
%
%\put(10,0){\circle{10}}
%
%
%\end{picture}
%}}%
%BeginExpansion
\setlength{\unitlength}{.05cm}
\begin{picture}(20,5)
\label{picc}

\put(0,0){\dashbox{0.5}(5,0){ }}
\put(15,0){\dashbox{0.5}(5,0){ }}

\thicklines

\put(10,0){\circle{10}}

\end{picture}
%
%EndExpansion
=%
%TCIMACRO{
%\TeXButton{identity}{\setlength{\unitlength}{.05cm}
%\begin{picture}(20,5)
%\label{picc}
%
%\put(0,0){\dashbox{0.5}(20,0){ }}
%
%
%\end{picture}
%}}%
%BeginExpansion
\setlength{\unitlength}{.05cm}
\begin{picture}(20,5)
\label{picc}

\put(0,0){\dashbox{0.5}(20,0){ }}

\end{picture}
%
%EndExpansion
+%
%TCIMACRO{
%\TeXButton{box}{\setlength{\unitlength}{.05cm}
%\begin{picture}(20,5)
%\label{picc}
%
%\put(0,0){\dashbox{0.5}(5,0){ }}
%
%\put(5,-5){\line(1,0){10}}
%\put(15,-5){\line(0,1){10}}
%\put(15,5){\line(-1,0){10}}
%\put(5,5){\line(0,-1){10}}
%
%\put(15,0){\dashbox{0.5}(5,0){ }}
%
%\end{picture}
%}}%
%BeginExpansion
\setlength{\unitlength}{.05cm}
\begin{picture}(20,5)
\label{picc}

\put(0,0){\dashbox{0.5}(5,0){ }}

\put(5,-5){\line(1,0){10}}
\put(15,-5){\line(0,1){10}}
\put(15,5){\line(-1,0){10}}
\put(5,5){\line(0,-1){10}}

\put(15,0){\dashbox{0.5}(5,0){ }}

\end{picture}
%
%EndExpansion
+%
%TCIMACRO{
%\TeXButton{box2}{\setlength{\unitlength}{.05cm}
%\begin{picture}(40,5)
%\label{picc}
%
%\put(0,0){\dashbox{0.5}(5,0){ }}
%
%\put(5,-5){\line(1,0){10}}
%\put(15,-5){\line(0,1){10}}
%\put(15,5){\line(-1,0){10}}
%\put(5,5){\line(0,-1){10}}
%
%\put(15,0){\dashbox{0.5}(5,0){ }}
%
%\put(20,-5){\line(1,0){10}}
%\put(30,-5){\line(0,1){10}}
%\put(30,5){\line(-1,0){10}}
%\put(20,5){\line(0,-1){10}}
%
%\put(30,0){\dashbox{0.5}(5,0){ }}
%
%\end{picture}
%} }%
%BeginExpansion
\setlength{\unitlength}{.05cm}
\begin{picture}(40,5)
\label{picc}

\put(0,0){\dashbox{0.5}(5,0){ }}

\put(5,-5){\line(1,0){10}}
\put(15,-5){\line(0,1){10}}
\put(15,5){\line(-1,0){10}}
\put(5,5){\line(0,-1){10}}

\put(15,0){\dashbox{0.5}(5,0){ }}

\put(20,-5){\line(1,0){10}}
\put(30,-5){\line(0,1){10}}
\put(30,5){\line(-1,0){10}}
\put(20,5){\line(0,-1){10}}

\put(30,0){\dashbox{0.5}(5,0){ }}

\end{picture}
%
%EndExpansion
+%
%TCIMACRO{
%\TeXButton{box3}{\setlength{\unitlength}{.05cm}
%\begin{picture}(50,5)
%\label{picc}
%
%\put(0,0){\dashbox{0.5}(5,0){ }}
%
%\put(5,-5){\line(1,0){10}}
%\put(15,-5){\line(0,1){10}}
%\put(15,5){\line(-1,0){10}}
%\put(5,5){\line(0,-1){10}}
%
%\put(15,0){\dashbox{0.5}(5,0){ }}
%
%\put(20,-5){\line(1,0){10}}
%\put(30,-5){\line(0,1){10}}
%\put(30,5){\line(-1,0){10}}
%\put(20,5){\line(0,-1){10}}
%
%\put(30,0){\dashbox{0.5}(5,0){ }}
%
%
%\put(35,-5){\line(1,0){10}}
%\put(45,-5){\line(0,1){10}}
%\put(45,5){\line(-1,0){10}}
%\put(35,5){\line(0,-1){10}}
%
%\put(45,0){\dashbox{0.5}(5,0){ }}
%
%
%\end{picture}
%}}%
%BeginExpansion
\setlength{\unitlength}{.05cm}
\begin{picture}(50,5)
\label{picc}

\put(0,0){\dashbox{0.5}(5,0){ }}

\put(5,-5){\line(1,0){10}}
\put(15,-5){\line(0,1){10}}
\put(15,5){\line(-1,0){10}}
\put(5,5){\line(0,-1){10}}

\put(15,0){\dashbox{0.5}(5,0){ }}

\put(20,-5){\line(1,0){10}}
\put(30,-5){\line(0,1){10}}
\put(30,5){\line(-1,0){10}}
\put(20,5){\line(0,-1){10}}

\put(30,0){\dashbox{0.5}(5,0){ }}

\put(35,-5){\line(1,0){10}}
\put(45,-5){\line(0,1){10}}
\put(45,5){\line(-1,0){10}}
\put(35,5){\line(0,-1){10}}

\put(45,0){\dashbox{0.5}(5,0){ }}

\end{picture}
%
%EndExpansion
+ $\ \ \cdots $

\bigskip
\end{center}

\noindent where now each box is the following sum over all effective
one-vertex contributions:

\bigskip

\begin{center}
%TCIMACRO{
%\TeXButton{boxbare}{\setlength{\unitlength}{.05cm}
%\begin{picture}(20,5)
%\label{picc}
%
%
%
%\put(5,-5){\line(1,0){10}}
%\put(15,-5){\line(0,1){10}}
%\put(15,5){\line(-1,0){10}}
%\put(5,5){\line(0,-1){10}}
%
%
%
%\end{picture}
%}}%
%BeginExpansion
\setlength{\unitlength}{.05cm}
\begin{picture}(20,5)
\label{picc}

\put(5,-5){\line(1,0){10}}
\put(15,-5){\line(0,1){10}}
\put(15,5){\line(-1,0){10}}
\put(5,5){\line(0,-1){10}}

\end{picture}
%
%EndExpansion
=%
%TCIMACRO{
%\TeXButton{one}{\setlength{\unitlength}{.05cm}
%\begin{picture}(10,5)
%\label{picc}
%
%
%
%\thicklines
%\put(5,0){\circle*{2}}
%
%\end{picture}
%}}%
%BeginExpansion
\setlength{\unitlength}{.05cm}
\begin{picture}(10,5)
\label{picc}

\thicklines
\put(5,0){\circle*{2}}

\end{picture}
%
%EndExpansion
+%
%TCIMACRO{
%\TeXButton{two contraction}{\setlength{\unitlength}{.05cm}
%\begin{picture}(10,5)
%\label{picc}
%
%\put(0,0){\dashbox{0.5}(10,0){ }}
%
%\thicklines
%\put(0,0){\circle*{2}}
%\put(10,0){\circle*{2}}
%
%\put(5,0){\oval(10,10)[t]}
%
%\end{picture}
%}}%
%BeginExpansion
\setlength{\unitlength}{.05cm}
\begin{picture}(10,5)
\label{picc}

\put(0,0){\dashbox{0.5}(10,0){ }}

\thicklines
\put(0,0){\circle*{2}}
\put(10,0){\circle*{2}}

\put(5,0){\oval(10,10)[t]}

\end{picture}
%
%EndExpansion
+%
%TCIMACRO{
%\TeXButton{three contraction}{\setlength{\unitlength}{.05cm}
%\begin{picture}(20,5)
%\label{picc}
%
%\put(0,0){\dashbox{0.5}(20,0){ }}
%
%\thicklines
%\put(0,0){\circle*{2}}
%\put(10,0){\circle*{2}}
%\put(20,0){\circle*{2}}
%
%\put(5,0){\oval(10,10)[t]}
%\put(15,0){\oval(10,10)[t]}
%
%\end{picture}
%}}%
%BeginExpansion
\setlength{\unitlength}{.05cm}
\begin{picture}(20,5)
\label{picc}

\put(0,0){\dashbox{0.5}(20,0){ }}

\thicklines
\put(0,0){\circle*{2}}
\put(10,0){\circle*{2}}
\put(20,0){\circle*{2}}

\put(5,0){\oval(10,10)[t]}
\put(15,0){\oval(10,10)[t]}

\end{picture}
%
%EndExpansion
+%
%TCIMACRO{
%\TeXButton{four contraction}{\setlength{\unitlength}{.05cm}
%\begin{picture}(30,5)
%\label{picc}
%
%\put(0,0){\dashbox{0.5}(30,0){ }}
%
%\thicklines
%\put(0,0){\circle*{2}}
%\put(10,0){\circle*{2}}
%\put(20,0){\circle*{2}}
%\put(30,0){\circle*{2}}
%
%\put(5,0){\oval(10,10)[t]}
%\put(15,0){\oval(10,10)[t]}
%\put(25,0){\oval(10,10)[t]}
%
%\end{picture}
%}}%
%BeginExpansion
\setlength{\unitlength}{.05cm}
\begin{picture}(30,5)
\label{picc}

\put(0,0){\dashbox{0.5}(30,0){ }}

\thicklines
\put(0,0){\circle*{2}}
\put(10,0){\circle*{2}}
\put(20,0){\circle*{2}}
\put(30,0){\circle*{2}}

\put(5,0){\oval(10,10)[t]}
\put(15,0){\oval(10,10)[t]}
\put(25,0){\oval(10,10)[t]}

\end{picture}
%
%EndExpansion
+%
%TCIMACRO{
%\TeXButton{five contraction}{\setlength{\unitlength}{.05cm}
%\begin{picture}(40,5)
%\label{picc}
%
%\put(0,0){\dashbox{0.5}(40,0){ }}
%
%\thicklines
%\put(0,0){\circle*{2}}
%\put(10,0){\circle*{2}}
%\put(20,0){\circle*{2}}
%\put(30,0){\circle*{2}}
%\put(40,0){\circle*{2}}
%
%\put(5,0){\oval(10,10)[t]}
%\put(15,0){\oval(10,10)[t]}
%\put(25,0){\oval(10,10)[t]}
%\put(35,0){\oval(10,10)[t]}
%
%\end{picture}
%}}%
%BeginExpansion
\setlength{\unitlength}{.05cm}
\begin{picture}(40,5)
\label{picc}

\put(0,0){\dashbox{0.5}(40,0){ }}

\thicklines
\put(0,0){\circle*{2}}
\put(10,0){\circle*{2}}
\put(20,0){\circle*{2}}
\put(30,0){\circle*{2}}
\put(40,0){\circle*{2}}

\put(5,0){\oval(10,10)[t]}
\put(15,0){\oval(10,10)[t]}
\put(25,0){\oval(10,10)[t]}
\put(35,0){\oval(10,10)[t]}

\end{picture}
%
%EndExpansion
+ $\cdots $

\bigskip
\end{center}

\noindent which is analogous to the expression of the self-energy in quantum
field theory:as a sum over irreducible terms. (As we have seen, one-vertex
contributions terminate at second order when there is no scattering: as this
is a form of cumulant expansion, the emission/absorption problem is
Gaussian, while allowing scattering means that we must have cumulant moments
to all orders!)

\bigskip

If the limit effective one-vertex label is $t$ then its weight is 
\begin{eqnarray*}
&&-i\tilde{E}_{00}\left( t\right) +\left( -i\right) ^{2}\kappa \tilde{E}%
_{01}\left( t\right) \tilde{E}_{10}\left( t\right) +\left( -i\right)
^{3}\kappa ^{2}\tilde{E}_{01}\left( t\right) \tilde{E}_{11}\left( t\right) 
\tilde{E}_{10}\left( t\right) +\cdots \\
&=&-i\tilde{E}_{00}\left( t\right) -\kappa \tilde{E}_{01}\left( t\right) 
\frac{1}{1+i\kappa E_{11}}\tilde{E}_{10}\left( t\right) \\
&\equiv &\left[ h_{1}^{\ast }\left( t\right) \right] ^{\alpha }G_{\alpha
\beta }\left[ h_{2}\left( t\right) \right] ^{\beta }
\end{eqnarray*}
where the geometric series can be summed since $\left\| \kappa
E_{11}\right\| <1$. We therefore see that 
\begin{eqnarray*}
&&\lim_{\lambda \rightarrow 0}\left\langle \phi _{1}\otimes \varepsilon
_{\lambda }\left( 1\right) \Phi _{R}|\,U_{t}\left( \lambda \right) \,\phi
_{2}\otimes \varepsilon _{\lambda }\left( 2\right) \Phi _{R}\right\rangle \\
&=&\left\langle \phi _{1}\otimes \varepsilon \left( f_{1}\otimes 1_{\left[
S_{1},T_{1}\right] }\right) \Phi |\,\left[ 1+\int_{0}^{t}G_{\alpha \beta
}dA^{\alpha \beta }\right] \,\phi _{2}\otimes \varepsilon \left(
f_{2}\otimes 1_{\left[ S_{2},T_{2}\right] }\right) \Phi \right\rangle .
\end{eqnarray*}

The QSDE then takes the form 
\begin{eqnarray*}
dU_{t} &=&G_{\alpha \beta }dA_{t}^{\alpha \beta }U_{t} \\
&=&\frac{1}{\gamma }\left( W-1\right) U_{t}\otimes dA_{t}^{11}+LU_{t}\otimes
dA_{t}^{10} \\
&&-L^{\dagger }WU_{t}\otimes dA_{t}^{01}-\left( \frac{1}{2}\gamma L^{\dagger
}L+iH\right) U_{t}\otimes dA_{t}^{00}
\end{eqnarray*}
with the coefficients $\left( W,L,H\right) $ are as before.

\section{Dynamical Evolutions}

Let $X$ be a bounded operator on the system state space $\frak{h}_{S}$. We
define its \textit{Heisenberg evolute} to be 
\begin{equation}
J_{t}^{\left( \lambda \right) }\left( X\right) :=U_{t}^{\left( \lambda
\right) \dagger }\,\left[ X\otimes 1_{R}\right] \,U_{t}^{\left( \lambda
\right) }.  \tag{9.1}
\end{equation}
In addition, what we term the \textit{co-evolute} is defined to be 
\begin{equation}
K_{t}^{\left( \lambda \right) }\left( X\right) :=U_{t}^{\left( \lambda
\right) }\,\left[ X\otimes 1_{R}\right] \,U_{t}^{\left( \lambda \right)
\dagger }.  \tag{9.2}
\end{equation}

We wish to study the limits of $J_{t}^{\left( \lambda \right) }$ and $%
K_{t}^{\left( \lambda \right) }$ as quantum processes taken relative to the
Fock vacuum state $\Phi \in \frak{h}_{R}$ for the Bose reservoir. To this
end, we note the developments 
\begin{eqnarray}
K_{t}^{\left( \lambda \right) }\left( X\right) &=&\sum_{n}\left( -1\right)
^{n}\int_{\Delta _{n}\left( t\right) }ds_{n}\cdots ds_{1}\,\mathcal{X}%
_{\Upsilon _{s_{n}}^{\left( \lambda \right) }}\circ \cdots \circ \mathcal{X}%
_{\Upsilon _{s_{1}}^{\left( \lambda \right) }}\left( X\otimes 1_{R}\right) ,
\notag \\
&&  \TCItag{9.3} \\
J_{t}^{\left( \lambda \right) }\left( X\right) &=&\sum_{n,\hat{n}}\left(
-i\right) ^{n+\hat{n}}\int_{\Delta _{n}\left( t\right) }ds_{n}\cdots
ds_{1}\,\int_{\Delta _{\hat{n}}\left( t\right) }dt_{\hat{n}}\cdots dt_{1}\, 
\notag \\
&&\times \Upsilon _{s_{1}}^{\left( \lambda \right) }\cdots \Upsilon
_{s_{n}}^{\left( \lambda \right) }\,\left[ X\otimes 1_{R}\right] \,\Upsilon
_{t_{\hat{n}}}^{\left( \lambda \right) }\cdots \Upsilon _{t_{1}}^{\left(
\lambda \right) },  \TCItag{9.4}
\end{eqnarray}
where $\mathcal{X}_{H}\left( .\right) :=\frac{1}{i}\left[ .,H\right] $.

\bigskip

We note that the co-evolution has the simpler form when iterated. The
evolution itself requires a separate expansion of the unitaries. (This
disparity is related to the proof of unitarity for quantum stochastic
processes in \cite{HP}, where the isometric property requires some work
while the co-isometric property is established immediately.) In fact, the
same inequalities as used to establish the convergence of $U_{t}^{\left(
\lambda \right) }$ suffice for the co-evolution: in both cases we have a
Picard iterated series. We remark that in \cite{AFL:Langevin} the
co-evolution only is treated for emission/absorption\ interactions.

We likewise have the expansion 
\begin{eqnarray}
&&\left\langle \phi _{1}\otimes W_{\lambda }\left( 1\right) \Phi
|\;J_{t}^{\left( \lambda \right) }\left( X\right) \;\phi _{2}\otimes
W_{\lambda }\left( 2\right) \Phi \right\rangle  \notag \\
&=&\sum_{n,\hat{n}}\left( -i\right) ^{n-\hat{n}}\int_{\Delta _{n}\left(
t\right) }ds_{n}\cdots ds_{1}\,\int_{\Delta _{\hat{n}}\left( t\right) }dt_{%
\hat{n}}\cdots dt_{1}\,  \notag \\
&&\times \left\langle \phi _{1}|\;\tilde{E}_{\alpha _{1}\beta _{1}}\left(
s_{1},\lambda \right) \dots \tilde{E}_{\alpha _{n}\beta _{n}}\left(
s_{n},\lambda \right) \,X\,\tilde{E}_{\mu _{\hat{n}}\nu _{\hat{n}}}\left( s_{%
\hat{n}},\lambda \right) \dots \tilde{E}_{\mu _{1}\nu _{1}}\left(
s_{1},\lambda \right) \;\phi _{2}\right\rangle  \notag \\
&&\times \left\langle \Phi |\,\left[ a_{s_{1}}^{+}\left( \lambda \right) %
\right] ^{\alpha _{1}}\left[ a_{s_{1}}^{-}\left( \lambda \right) \right]
^{\beta _{1}}\cdots \left[ a_{s_{n}}^{+}\left( \lambda \right) \right]
^{\alpha _{n}}\left[ a_{s_{n}}^{-}\left( \lambda \right) \right] ^{\beta
_{n}}\right.  \notag \\
&&\left. \left[ a_{t_{\hat{n}}}^{+}\left( \lambda \right) \right] ^{\mu _{%
\hat{n}}}\left[ a_{t_{\hat{n}}}^{-}\left( \lambda \right) \right] ^{\nu _{%
\hat{n}}}\dots \left[ a_{t_{1}}^{+}\left( \lambda \right) \right] ^{\mu _{1}}%
\left[ a_{t_{1}}^{-}\left( \lambda \right) \right] ^{\nu _{1}}\,\Phi
\right\rangle .  \TCItag{9.5}
\end{eqnarray}
The vacuum average of the reservoir operators can be expressed as a sum of
products of two-point functions with each summand representable as a
partition of $n+\hat{n}$ vertices. Our strategy is similar to before. We
shall use diagrams to describe the individual contributions, and attempt to
obtain a uniform estimate. The Heisenberg diagrams are a more involved than
last time due to the scattering, however, the general idea goes through
again.

Let us consider an arbitrary Heisenberg diagram. If we considered only the $%
t-t$ contractions and ignored everything else then we would have a partition
of the $n$ $t-$variables, let's say with occupation numbers $\mathbf{n}%
=\left( n_{j}\right) $. Likewise, if we looked at only the $s-s$
contractions then we have a partition of the $n^{\prime }$ $s-$variables,
say with occupation numbers $\mathbf{n}^{\prime }=\left( n_{j}^{\prime
}\right) $. At this stage we can then take the $s-t$ contractions into
account. The diagram below shows a quartet of $s$ variables joined to a
triple of $t$ variables.

\bigskip

\begin{center}
%TCIMACRO{
%\TeXButton{mixed}{\setlength{\unitlength}{.10cm}
%\begin{picture}(50,15)
%\label{picc}
%
%
%\thicklines
%
%
%\put(0,5){\circle*{2}}
%\put(5,5){\circle*{2}}
%\put(10,5){\circle*{2}}
%\put(15,5){\circle*{2}}
%
%\put(35,5){\circle*{2}}
%\put(40,5){\circle*{2}}
%\put(45,5){\circle*{2}}
%
%\put(2.5,5){\oval(5,5)[t]}
%
%\put(7.5,5){\oval(5,5)[t]}
%\put(12.5,5){\oval(5,5)[t]}
%\put(37.5,5){\oval(5,5)[t]}
%\put(42.5,5){\oval(5,5)[t]}
%
%\put(25,5){\oval(20,15)[t]}
%\put(0,1){$s$}
%\put(5,1){$s$}
%\put(10,1){$s$}
%\put(15,1){$s$}
%\put(24,6){$X$}
%\put(35,1){$t$}
%\put(40,1){$t$}
%\put(45,1){$t$}
%\end{picture}
%}}%
%BeginExpansion
\setlength{\unitlength}{.10cm}
\begin{picture}(50,15)
\label{picc}

\thicklines

\put(0,5){\circle*{2}}
\put(5,5){\circle*{2}}
\put(10,5){\circle*{2}}
\put(15,5){\circle*{2}}

\put(35,5){\circle*{2}}
\put(40,5){\circle*{2}}
\put(45,5){\circle*{2}}

\put(2.5,5){\oval(5,5)[t]}

\put(7.5,5){\oval(5,5)[t]}
\put(12.5,5){\oval(5,5)[t]}
\put(37.5,5){\oval(5,5)[t]}
\put(42.5,5){\oval(5,5)[t]}

\put(25,5){\oval(20,15)[t]}
\put(0,1){$s$}
\put(5,1){$s$}
\put(10,1){$s$}
\put(15,1){$s$}
\put(24,6){$X$}
\put(35,1){$t$}
\put(40,1){$t$}
\put(45,1){$t$}
\end{picture}
%
%EndExpansion

Figure 5
\end{center}

Let $l_{jk}$ be the number of $s-t$ contractions joining a part of $j$ $s$'s
to a part of $k$ $t$'s: here we use an obvious abuse of terminology, as
technically they are all in the same part! We also introduce the occupation
numbers $\mathbf{l}=\left( l_{j}\right) $, $\mathbf{l}^{\prime }=\left(
l_{j}^{\prime }\right) $ where $l_{k}=\sum_{j}l_{jk}$ and $l_{j}^{\prime
}=\sum_{k}l_{jk}$. (When no scattering was present, we only had the
possibility that $l_{11}$, previously denoted as $l$, could be non-zero.) It
is convenient to introduce the occupation numbers $\mathbf{m}=\left(
m_{j}\right) $ and $\mathbf{m}^{\prime }=\left( m_{j}^{\prime }\right) $
where $m_{j}=n_{j}-l_{j}$ and $m_{j}^{\prime }=n_{j}^{\prime }-l_{j}^{\prime
}$. Here $m_{j}$ counts the number of parts of $t$-variables of size $j$
having no elements contracted with an $s$-variable.

The procedure adopted in the last chapter is now repeated. We consider
equivalence classes of Heisenberg diagrams leading to the same set of
sequences $\mathbf{n},\mathbf{n}^{\prime },\mathbf{l},\mathbf{l}^{\prime }$,
or equivalently $\mathbf{m},\mathbf{m}^{\prime },\mathbf{l},\mathbf{l}%
^{\prime }$ as above. We can choose a basic Heisenberg diagram as the
representative of each class, and there will be permutations $\rho \in \frak{%
S}_{\mathbf{n}}^{0}$ and $\rho ^{\prime }\in \frak{S}_{\mathbf{n}^{\prime
}}^{0}$ of the $t$ and $s$ variables respectively which will allow us to
reorganize the basic Heisenberg diagram into any other element of the the
class. (We omit the explicit choice of basic of Heisenberg diagram and leave
its specification to the reader as an exercise.)

Now for each diagram in a given class there will then be chronologically
ordered blocks of sizes $m_{1},m_{2},\cdots ,m_{1}^{\prime }$

$,m_{2}^{\prime },\cdots ,l_{1},l_{2},\cdots $ and by the type of argument
encountered before we arrive at the following upper bound for the sum of
absolute values for all the diagrams 
\begin{eqnarray*}
&&\sum_{\mathbf{m},\mathbf{m}^{\prime },\mathbf{l}}C_{11}^{E\left( \mathbf{m}%
+\mathbf{m}^{\prime }+\mathbf{l}+\mathbf{l}^{\prime }\right) -2N\left( 
\mathbf{m}+\mathbf{m}^{\prime }+\mathbf{l}+\mathbf{l}^{\prime }\right)
+m_{1}+m_{1}^{\prime }+l_{1}+l_{2}} \\
&&\times C^{2N\left( \mathbf{m}+\mathbf{m}^{\prime }+\mathbf{l}+\mathbf{l}%
^{\prime }\right) -\left( m_{1}+m_{1}^{\prime }+l_{1}+l_{2}\right) } \\
&&\times \frac{\left( t\vee 1\right) ^{N\left( \mathbf{m}+\mathbf{m}^{\prime
}+\mathbf{l}\right) }}{\left( m_{1}!m_{2}!\cdots \right) \left(
m_{1}^{\prime }!m_{2}^{\prime }!\cdots \right) \left( l_{1}!l_{2}!\cdots
\right) } \\
&&\times K^{E\left( \mathbf{m}+\mathbf{m}^{\prime }+\mathbf{l}+\mathbf{l}%
^{\prime }\right) -N\left( \mathbf{m}+\mathbf{m}^{\prime }+\mathbf{l}+%
\mathbf{l}^{\prime }\right) }\gamma ^{N\left( \mathbf{l}\right) }.
\end{eqnarray*}
Here we add sequences of occupation numbers componentwise, ie $\mathbf{m}+%
\mathbf{m}^{\prime }$ is $\left( m_{j}+m_{j}^{\prime }\right) $, etc., and
we note that $N\left( \mathbf{l}\right) =N\left( \mathbf{l}^{\prime }\right) 
$. Recalling the constants $A$ and $B$ from before, and introducing $%
B^{\prime }=\frac{1}{2}\ln \left( t\vee 1\right) +\ln \left( C^{2}\vee
1\right) +\ln \left( C_{11}^{-2}\vee 1\right) +\ln \left( K^{-1}\vee
1\right) +\frac{1}{2}\ln \left( \gamma \right) $, we sum the series to get
the upperbound 
\begin{equation*}
\exp \left\{ 2\frac{e^{A+B}}{1-e^{A}}+\frac{e^{2A+2B^{\prime }}}{1-e^{2A}}%
\right\}
\end{equation*}
which is again convergent as $e^{A}<1$.

We now wish to determine the limit $\lambda \rightarrow 0$. Once again, only
diagrams having time consecutive $s-s$ and $t-t$ contractions, as well as
non-crossing $s-t$ contractions, are going to contribute to the limit. The
presence of scattering now means that we have more diagrams, however, we can
reduce this using the effective vertex method and, once again we can arrive
at a simple recursive formula. This time, we have 
\begin{eqnarray*}
\left( 
%TCIMACRO{
%\TeXButton{vacuum}{\setlength{\unitlength}{.05cm}
%\begin{picture}(20,5)
%\label{picc}
%
%\put(0,3){\dashbox{0.5}(5,0){ }}
%\put(15,3){\dashbox{0.5}(5,0){ }}
%
%\thicklines
%
%\put(10,3){\circle{10}}
%
%
%\end{picture}
%}}%
%BeginExpansion
\setlength{\unitlength}{.05cm}
\begin{picture}(20,5)
\label{picc}

\put(0,3){\dashbox{0.5}(5,0){ }}
\put(15,3){\dashbox{0.5}(5,0){ }}

\thicklines

\put(10,3){\circle{10}}

\end{picture}
%
%EndExpansion
X%
%TCIMACRO{
%\TeXButton{vacuum}{\setlength{\unitlength}{.05cm}
%\begin{picture}(20,5)
%\label{picc}
%
%\put(0,3){\dashbox{0.5}(5,0){ }}
%\put(15,3){\dashbox{0.5}(5,0){ }}
%
%\thicklines
%
%\put(10,3){\circle{10}}
%
%
%\end{picture}
%}}%
%BeginExpansion
\setlength{\unitlength}{.05cm}
\begin{picture}(20,5)
\label{picc}

\put(0,3){\dashbox{0.5}(5,0){ }}
\put(15,3){\dashbox{0.5}(5,0){ }}

\thicklines

\put(10,3){\circle{10}}

\end{picture}
%
%EndExpansion
\right) &=&\left( 
%TCIMACRO{
%\TeXButton{identity}{\setlength{\unitlength}{.05cm}
%\begin{picture}(20,5)
%\label{picc}
%
%\put(0,3){\dashbox{0.5}(20,0){ }}
%
%
%\end{picture}
%}}%
%BeginExpansion
\setlength{\unitlength}{.05cm}
\begin{picture}(20,5)
\label{picc}

\put(0,3){\dashbox{0.5}(20,0){ }}

\end{picture}
%
%EndExpansion
X%
%TCIMACRO{
%\TeXButton{identity}{\setlength{\unitlength}{.05cm}
%\begin{picture}(20,5)
%\label{picc}
%
%\put(0,3){\dashbox{0.5}(20,0){ }}
%
%
%\end{picture}
%}}%
%BeginExpansion
\setlength{\unitlength}{.05cm}
\begin{picture}(20,5)
\label{picc}

\put(0,3){\dashbox{0.5}(20,0){ }}

\end{picture}
%
%EndExpansion
\right) \\
&& \\
&&+\left( 
%TCIMACRO{
%\TeXButton{vacuum}{\setlength{\unitlength}{.05cm}
%\begin{picture}(20,5)
%\label{picc}
%
%\put(0,3){\dashbox{0.5}(5,0){ }}
%\put(15,3){\dashbox{0.5}(5,0){ }}
%
%\thicklines
%
%\put(10,3){\circle{10}}
%
%
%\end{picture}
%}}%
%BeginExpansion
\setlength{\unitlength}{.05cm}
\begin{picture}(20,5)
\label{picc}

\put(0,3){\dashbox{0.5}(5,0){ }}
\put(15,3){\dashbox{0.5}(5,0){ }}

\thicklines

\put(10,3){\circle{10}}

\end{picture}
%
%EndExpansion
X%
%TCIMACRO{
%\TeXButton{box}{\setlength{\unitlength}{.05cm}
%\begin{picture}(40,10)
%\label{picc}
%
%\put(0,3){\dashbox{0.5}(5,0){ }}
%
%\put(5,-2){\line(1,0){10}}
%\put(15,-2){\line(0,1){10}}
%\put(15,8){\line(-1,0){10}}
%\put(5,8){\line(0,-1){10}}
%
%\put(15,3){\dashbox{0.5}(5,0){ }}
%
%\put(30,3){\dashbox{0.5}(5,0){ }}
%
%\thicklines
%
%\put(25,3){\circle{10}}
%
%\end{picture}
%}}%
%BeginExpansion
\setlength{\unitlength}{.05cm}
\begin{picture}(40,10)
\label{picc}

\put(0,3){\dashbox{0.5}(5,0){ }}

\put(5,-2){\line(1,0){10}}
\put(15,-2){\line(0,1){10}}
\put(15,8){\line(-1,0){10}}
\put(5,8){\line(0,-1){10}}

\put(15,3){\dashbox{0.5}(5,0){ }}

\put(30,3){\dashbox{0.5}(5,0){ }}

\thicklines

\put(25,3){\circle{10}}

\end{picture}
%
%EndExpansion
\right) \\
&& \\
&&+\left( 
%TCIMACRO{
%\TeXButton{box}{\setlength{\unitlength}{.05cm}
%\begin{picture}(35,10)
%\label{picc}
%
%\put(0,3){\dashbox{0.5}(5,0){ }}
%
%\put(20,-2){\line(1,0){10}}
%\put(30,-2){\line(0,1){10}}
%\put(30,8){\line(-1,0){10}}
%\put(20,8){\line(0,-1){10}}
%
%\put(15,3){\dashbox{0.5}(5,0){ }}
%
%\put(30,3){\dashbox{0.5}(5,0){ }}
%
%\thicklines
%
%\put(10,3){\circle{10}}
%
%\end{picture}
%}}%
%BeginExpansion
\setlength{\unitlength}{.05cm}
\begin{picture}(35,10)
\label{picc}

\put(0,3){\dashbox{0.5}(5,0){ }}

\put(20,-2){\line(1,0){10}}
\put(30,-2){\line(0,1){10}}
\put(30,8){\line(-1,0){10}}
\put(20,8){\line(0,-1){10}}

\put(15,3){\dashbox{0.5}(5,0){ }}

\put(30,3){\dashbox{0.5}(5,0){ }}

\thicklines

\put(10,3){\circle{10}}

\end{picture}
%
%EndExpansion
X%
%TCIMACRO{
%\TeXButton{vacuum}{\setlength{\unitlength}{.05cm}
%\begin{picture}(20,5)
%\label{picc}
%
%\put(0,3){\dashbox{0.5}(5,0){ }}
%\put(15,3){\dashbox{0.5}(5,0){ }}
%
%\thicklines
%
%\put(10,3){\circle{10}}
%
%
%\end{picture}
%}}%
%BeginExpansion
\setlength{\unitlength}{.05cm}
\begin{picture}(20,5)
\label{picc}

\put(0,3){\dashbox{0.5}(5,0){ }}
\put(15,3){\dashbox{0.5}(5,0){ }}

\thicklines

\put(10,3){\circle{10}}

\end{picture}
%
%EndExpansion
\right) \\
&& \\
&&+\left( 
%TCIMACRO{
%\TeXButton{identity}{\setlength{\unitlength}{.05cm}
%\begin{picture}(115,15)
%\label{picc}
%
%\put(0,0){\dashbox{0.5}(10,0){ }}
%\put(20,0){\dashbox{0.5}(10,0){ }}
%\put(40,0){\dashbox{0.5}(10,0){ }}
%\put(65,0){\dashbox{0.5}(10,0){ }}
%\put(85,0){\dashbox{0.5}(10,0){ }}
%\put(105,0){\dashbox{0.5}(10,0){ }}
%\put(55,-2){$X$}
%\thicklines
%\put(40,0){\circle*{2}}
%\put(75,0){\circle*{2}}
%
%\put(30,-5){\line(1,0){10}}
%\put(40,-5){\line(0,1){10}}
%\put(40,5){\line(-1,0){10}}
%\put(30,5){\line(0,-1){10}}
%
%
%\put(75,-5){\line(1,0){10}}
%\put(85,-5){\line(0,1){10}}
%\put(85,5){\line(-1,0){10}}
%\put(75,5){\line(0,-1){10}}
%
%\put(15,0){\circle{10}}
%\put(100,0){\circle{10}}
%\put(57.5,0){\oval(35,25)[t]}
%\end{picture}
%}}%
%BeginExpansion
\setlength{\unitlength}{.05cm}
\begin{picture}(115,15)
\label{picc}

\put(0,0){\dashbox{0.5}(10,0){ }}
\put(20,0){\dashbox{0.5}(10,0){ }}
\put(40,0){\dashbox{0.5}(10,0){ }}
\put(65,0){\dashbox{0.5}(10,0){ }}
\put(85,0){\dashbox{0.5}(10,0){ }}
\put(105,0){\dashbox{0.5}(10,0){ }}
\put(55,-2){$X$}
\thicklines
\put(40,0){\circle*{2}}
\put(75,0){\circle*{2}}

\put(30,-5){\line(1,0){10}}
\put(40,-5){\line(0,1){10}}
\put(40,5){\line(-1,0){10}}
\put(30,5){\line(0,-1){10}}

\put(75,-5){\line(1,0){10}}
\put(85,-5){\line(0,1){10}}
\put(85,5){\line(-1,0){10}}
\put(75,5){\line(0,-1){10}}

\put(15,0){\circle{10}}
\put(100,0){\circle{10}}
\put(57.5,0){\oval(35,25)[t]}
\end{picture}
%
%EndExpansion
\right)
\end{eqnarray*}

Here we meet new effective vertices in the final diagram. On the right we
have

\begin{center}
%TCIMACRO{
%\TeXButton{boxbare}{\setlength{\unitlength}{.05cm}
%\begin{picture}(20,5)
%\label{picc}
%
%\thicklines
%
%\put(5,0){\circle*{2}}
%\put(0,0){\oval(10,10)[tr]}
%
%\put(5,-5){\line(1,0){10}}
%\put(15,-5){\line(0,1){10}}
%\put(15,5){\line(-1,0){10}}
%\put(5,5){\line(0,-1){10}}
%
%
%
%\end{picture}
%}}%
%BeginExpansion
\setlength{\unitlength}{.05cm}
\begin{picture}(20,5)
\label{picc}

\thicklines

\put(5,0){\circle*{2}}
\put(0,0){\oval(10,10)[tr]}

\put(5,-5){\line(1,0){10}}
\put(15,-5){\line(0,1){10}}
\put(15,5){\line(-1,0){10}}
\put(5,5){\line(0,-1){10}}

\end{picture}
%
%EndExpansion
=%
%TCIMACRO{
%\TeXButton{one}{\setlength{\unitlength}{.05cm}
%\begin{picture}(10,5)
%\label{picc}
%
%
%
%\thicklines
%
%\put(0,0){\oval(10,10)[tr]}
%\put(5,0){\circle*{2}}
%
%\end{picture}
%}}%
%BeginExpansion
\setlength{\unitlength}{.05cm}
\begin{picture}(10,5)
\label{picc}

\thicklines

\put(0,0){\oval(10,10)[tr]}
\put(5,0){\circle*{2}}

\end{picture}
%
%EndExpansion
+\quad 
%TCIMACRO{
%\TeXButton{two contraction}{\setlength{\unitlength}{.05cm}
%\begin{picture}(10,5)
%\label{picc}
%
%\put(0,0){\dashbox{0.5}(10,0){ }}
%
%\thicklines
%
%\put(-5,0){\oval(10,10)[tr]}
%\put(0,0){\circle*{2}}
%\put(10,0){\circle*{2}}
%
%\put(5,0){\oval(10,10)[t]}
%
%\end{picture}
%}}%
%BeginExpansion
\setlength{\unitlength}{.05cm}
\begin{picture}(10,5)
\label{picc}

\put(0,0){\dashbox{0.5}(10,0){ }}

\thicklines

\put(-5,0){\oval(10,10)[tr]}
\put(0,0){\circle*{2}}
\put(10,0){\circle*{2}}

\put(5,0){\oval(10,10)[t]}

\end{picture}
%
%EndExpansion
+\quad 
%TCIMACRO{
%\TeXButton{three contraction}{\setlength{\unitlength}{.05cm}
%\begin{picture}(20,5)
%\label{picc}
%
%\put(0,0){\dashbox{0.5}(20,0){ }}
%
%\thicklines
%
%\put(-5,0){\oval(10,10)[tr]}
%\put(0,0){\circle*{2}}
%\put(10,0){\circle*{2}}
%\put(20,0){\circle*{2}}
%
%\put(5,0){\oval(10,10)[t]}
%\put(15,0){\oval(10,10)[t]}
%
%\end{picture}
%}}%
%BeginExpansion
\setlength{\unitlength}{.05cm}
\begin{picture}(20,5)
\label{picc}

\put(0,0){\dashbox{0.5}(20,0){ }}

\thicklines

\put(-5,0){\oval(10,10)[tr]}
\put(0,0){\circle*{2}}
\put(10,0){\circle*{2}}
\put(20,0){\circle*{2}}

\put(5,0){\oval(10,10)[t]}
\put(15,0){\oval(10,10)[t]}

\end{picture}
%
%EndExpansion
+\quad 
%TCIMACRO{
%\TeXButton{four contraction}{\setlength{\unitlength}{.05cm}
%\begin{picture}(30,5)
%\label{picc}
%
%\put(0,0){\dashbox{0.5}(30,0){ }}
%
%\thicklines
%
%\put(-5,0){\oval(10,10)[tr]}
%\put(0,0){\circle*{2}}
%\put(10,0){\circle*{2}}
%\put(20,0){\circle*{2}}
%\put(30,0){\circle*{2}}
%
%\put(5,0){\oval(10,10)[t]}
%\put(15,0){\oval(10,10)[t]}
%\put(25,0){\oval(10,10)[t]}
%
%\end{picture}
%}}%
%BeginExpansion
\setlength{\unitlength}{.05cm}
\begin{picture}(30,5)
\label{picc}

\put(0,0){\dashbox{0.5}(30,0){ }}

\thicklines

\put(-5,0){\oval(10,10)[tr]}
\put(0,0){\circle*{2}}
\put(10,0){\circle*{2}}
\put(20,0){\circle*{2}}
\put(30,0){\circle*{2}}

\put(5,0){\oval(10,10)[t]}
\put(15,0){\oval(10,10)[t]}
\put(25,0){\oval(10,10)[t]}

\end{picture}
%
%EndExpansion
+\quad 
%TCIMACRO{
%\TeXButton{five contraction}{\setlength{\unitlength}{.05cm}
%\begin{picture}(40,5)
%\label{picc}
%
%\put(0,0){\dashbox{0.5}(40,0){ }}
%
%\thicklines
%
%\put(-5,0){\oval(10,10)[tr]}
%\put(0,0){\circle*{2}}
%\put(10,0){\circle*{2}}
%\put(20,0){\circle*{2}}
%\put(30,0){\circle*{2}}
%\put(40,0){\circle*{2}}
%
%\put(5,0){\oval(10,10)[t]}
%\put(15,0){\oval(10,10)[t]}
%\put(25,0){\oval(10,10)[t]}
%\put(35,0){\oval(10,10)[t]}
%
%\end{picture}
%}}%
%BeginExpansion
\setlength{\unitlength}{.05cm}
\begin{picture}(40,5)
\label{picc}

\put(0,0){\dashbox{0.5}(40,0){ }}

\thicklines

\put(-5,0){\oval(10,10)[tr]}
\put(0,0){\circle*{2}}
\put(10,0){\circle*{2}}
\put(20,0){\circle*{2}}
\put(30,0){\circle*{2}}
\put(40,0){\circle*{2}}

\put(5,0){\oval(10,10)[t]}
\put(15,0){\oval(10,10)[t]}
\put(25,0){\oval(10,10)[t]}
\put(35,0){\oval(10,10)[t]}

\end{picture}
%
%EndExpansion
+ $\cdots $
\end{center}

\bigskip

\noindent which for vertex time $t$ corresponds to the operator weigth 
\begin{eqnarray*}
&&-i\tilde{E}_{10}\left( t\right) +\left( -i\right) ^{2}\kappa \tilde{E}%
_{11}\left( t\right) \tilde{E}_{10}\left( t\right) +\left( -i\right)
^{3}\kappa ^{2}\tilde{E}_{11}\left( t\right) \tilde{E}_{11}\left( t\right) 
\tilde{E}_{10}\left( t\right) +\cdots \\
&=&-i\frac{1}{1+i\kappa E_{11}}\tilde{E}_{10}\left( t\right) \\
&\equiv &G_{1\beta }\left[ h_{2}\left( t\right) \right] ^{\beta }.
\end{eqnarray*}
While on the left we have

\begin{center}
%TCIMACRO{
%\TeXButton{boxbare}{\setlength{\unitlength}{.05cm}
%\begin{picture}(20,5)
%\label{picc}
%
%\thicklines
%
%\put(5,-5){\line(1,0){10}}
%\put(15,-5){\line(0,1){10}}
%\put(15,5){\line(-1,0){10}}
%\put(5,5){\line(0,-1){10}}
%
%\put(15,0){\circle*{2}}
%
%\put(20,0){\oval(10,10)[tl]}
%
%
%\end{picture}
%}}%
%BeginExpansion
\setlength{\unitlength}{.05cm}
\begin{picture}(20,5)
\label{picc}

\thicklines

\put(5,-5){\line(1,0){10}}
\put(15,-5){\line(0,1){10}}
\put(15,5){\line(-1,0){10}}
\put(5,5){\line(0,-1){10}}

\put(15,0){\circle*{2}}

\put(20,0){\oval(10,10)[tl]}

\end{picture}
%
%EndExpansion
=%
%TCIMACRO{
%\TeXButton{one}{\setlength{\unitlength}{.05cm}
%\begin{picture}(15,5)
%\label{picc}
%
%
%
%\thicklines
%\put(5,0){\circle*{2}}
%\put(10,0){\oval(10,10)[tl]}
%
%\end{picture}
%}}%
%BeginExpansion
\setlength{\unitlength}{.05cm}
\begin{picture}(15,5)
\label{picc}

\thicklines
\put(5,0){\circle*{2}}
\put(10,0){\oval(10,10)[tl]}

\end{picture}
%
%EndExpansion
+%
%TCIMACRO{
%\TeXButton{two contraction}{\setlength{\unitlength}{.05cm}
%\begin{picture}(15,5)
%\label{picc}
%
%\put(0,0){\dashbox{0.5}(10,0){ }}
%
%\thicklines
%\put(0,0){\circle*{2}}
%\put(10,0){\circle*{2}}
%
%\put(5,0){\oval(10,10)[t]}
%
%
%\put(15,0){\oval(10,10)[tl]}
%
%\end{picture}
%}}%
%BeginExpansion
\setlength{\unitlength}{.05cm}
\begin{picture}(15,5)
\label{picc}

\put(0,0){\dashbox{0.5}(10,0){ }}

\thicklines
\put(0,0){\circle*{2}}
\put(10,0){\circle*{2}}

\put(5,0){\oval(10,10)[t]}

\put(15,0){\oval(10,10)[tl]}

\end{picture}
%
%EndExpansion
+%
%TCIMACRO{
%\TeXButton{three contraction}{\setlength{\unitlength}{.05cm}
%\begin{picture}(25,5)
%\label{picc}
%
%\put(0,0){\dashbox{0.5}(20,0){ }}
%
%\thicklines
%\put(0,0){\circle*{2}}
%\put(10,0){\circle*{2}}
%\put(20,0){\circle*{2}}
%
%\put(5,0){\oval(10,10)[t]}
%\put(15,0){\oval(10,10)[t]}
%\put(25,0){\oval(10,10)[tl]}
%
%\end{picture}
%}}%
%BeginExpansion
\setlength{\unitlength}{.05cm}
\begin{picture}(25,5)
\label{picc}

\put(0,0){\dashbox{0.5}(20,0){ }}

\thicklines
\put(0,0){\circle*{2}}
\put(10,0){\circle*{2}}
\put(20,0){\circle*{2}}

\put(5,0){\oval(10,10)[t]}
\put(15,0){\oval(10,10)[t]}
\put(25,0){\oval(10,10)[tl]}

\end{picture}
%
%EndExpansion
+%
%TCIMACRO{
%\TeXButton{four contraction}{\setlength{\unitlength}{.05cm}
%\begin{picture}(35,5)
%\label{picc}
%
%\put(0,0){\dashbox{0.5}(30,0){ }}
%
%\thicklines
%\put(0,0){\circle*{2}}
%\put(10,0){\circle*{2}}
%\put(20,0){\circle*{2}}
%\put(30,0){\circle*{2}}
%
%\put(5,0){\oval(10,10)[t]}
%\put(15,0){\oval(10,10)[t]}
%\put(25,0){\oval(10,10)[t]}
%\put(35,0){\oval(10,10)[tl]}
%
%\end{picture}
%}}%
%BeginExpansion
\setlength{\unitlength}{.05cm}
\begin{picture}(35,5)
\label{picc}

\put(0,0){\dashbox{0.5}(30,0){ }}

\thicklines
\put(0,0){\circle*{2}}
\put(10,0){\circle*{2}}
\put(20,0){\circle*{2}}
\put(30,0){\circle*{2}}

\put(5,0){\oval(10,10)[t]}
\put(15,0){\oval(10,10)[t]}
\put(25,0){\oval(10,10)[t]}
\put(35,0){\oval(10,10)[tl]}

\end{picture}
%
%EndExpansion
+%
%TCIMACRO{
%\TeXButton{five contraction}{\setlength{\unitlength}{.05cm}
%\begin{picture}(45,5)
%\label{picc}
%
%\put(0,0){\dashbox{0.5}(40,0){ }}
%
%\thicklines
%\put(0,0){\circle*{2}}
%\put(10,0){\circle*{2}}
%\put(20,0){\circle*{2}}
%\put(30,0){\circle*{2}}
%\put(40,0){\circle*{2}}
%
%\put(5,0){\oval(10,10)[t]}
%\put(15,0){\oval(10,10)[t]}
%\put(25,0){\oval(10,10)[t]}
%\put(35,0){\oval(10,10)[t]}
%\put(45,0){\oval(10,10)[tl]}
%
%\end{picture}
%}}%
%BeginExpansion
\setlength{\unitlength}{.05cm}
\begin{picture}(45,5)
\label{picc}

\put(0,0){\dashbox{0.5}(40,0){ }}

\thicklines
\put(0,0){\circle*{2}}
\put(10,0){\circle*{2}}
\put(20,0){\circle*{2}}
\put(30,0){\circle*{2}}
\put(40,0){\circle*{2}}

\put(5,0){\oval(10,10)[t]}
\put(15,0){\oval(10,10)[t]}
\put(25,0){\oval(10,10)[t]}
\put(35,0){\oval(10,10)[t]}
\put(45,0){\oval(10,10)[tl]}

\end{picture}
%
%EndExpansion
+ $\cdots $

\bigskip
\end{center}

\noindent which has the weight 
\begin{eqnarray*}
&&i\tilde{E}_{01}\left( t\right) +i^{2}\kappa ^{\ast }\tilde{E}_{01}\left(
t\right) \tilde{E}_{11}\left( t\right) +i^{3}\left( \kappa ^{\ast }\right)
^{2}\tilde{E}_{01}\left( t\right) \tilde{E}_{11}\left( t\right) \tilde{E}%
_{11}\left( t\right) +\cdots \\
&=&i\tilde{E}_{01}\left( t\right) \frac{1}{1-i\kappa ^{\ast }E_{11}} \\
&\equiv &\left[ h_{1}\left( t\right) ^{\ast }\right] ^{\alpha }G_{1\alpha
}^{\dag }.
\end{eqnarray*}
The recursion relation here is 
\begin{gather*}
\left\langle \phi _{1}\otimes \varepsilon \left( 1\right) |U_{t}^{\dag } 
\left[ X\otimes 1\right] U_{t}\phi _{2}\otimes \varepsilon \left( 2\right)
\right\rangle =\left\langle \phi _{1}\otimes \varepsilon \left( 1\right) |%
\left[ X\otimes 1\right] \phi _{2}\otimes \varepsilon \left( 2\right)
\right\rangle \\
+\int_{\Delta _{2}\left( t\right) }\left\langle \phi _{1}\otimes \varepsilon
\left( 1\right) |U_{t_{2}}^{\dag }\left[ X\otimes 1\right] \left( G_{\alpha
\beta }\left[ h_{1}\left( t_{1}\right) ^{\ast }\right] ^{\alpha }\left[
h_{2}\left( t_{1}\right) \right] ^{\beta }\right) dU_{t_{1}}\phi _{2}\otimes
\varepsilon \left( 2\right) \right\rangle \\
+\int_{\Delta _{2}\left( t\right) }\left\langle \phi _{1}\otimes \varepsilon
\left( 1\right) |dU_{t_{1}}^{\dag }\left( G_{\beta \alpha }^{\dag }\left[
h_{1}\left( t_{1}\right) ^{\ast }\right] ^{\alpha }\left[ h_{2}\left(
t_{1}\right) \right] ^{\beta }\right) \left[ X\otimes 1\right] U_{t_{2}}\phi
_{2}\otimes \varepsilon \left( 2\right) \right\rangle \\
+\int_{0}^{t}dt_{1}\left\langle \phi _{1}\otimes \varepsilon \left( 1\right)
|U_{t_{1}}^{\dag }\left[ \left[ h_{1}\left( t_{1}\right) ^{\ast }\right]
^{\alpha }G_{1\alpha }^{\dag }XG_{1\beta }\left[ h_{2}\left( t_{1}\right) %
\right] ^{\beta }\otimes 1\right] U_{t_{1}}\phi _{2}\otimes \varepsilon
\left( 2\right) \right\rangle
\end{gather*}
\begin{eqnarray*}
&=&\phi _{1}\otimes \varepsilon \left( 1\right) |\left\{
1+\int_{0}^{t}U_{s}^{\dag }\left[ \left( XG_{\alpha \beta }+G_{\beta \alpha
}^{\dag }X+G_{1\alpha }^{\dag }XG_{1\beta }\right) \otimes dA_{s}^{\alpha
\beta }\right] \otimes dA_{s}^{\alpha \beta }U_{s}\right\} \\
&\equiv &\left\langle \phi _{1}\otimes \varepsilon \left( 1\right) |\left\{
1+\int_{0}^{t}U_{s}^{\dag }\mathcal{L}_{\alpha \beta }\left( X\right)
U_{s}dA_{s}^{\alpha \beta }\right\} \phi _{2}\otimes \varepsilon \left(
2\right) \right\rangle ,
\end{eqnarray*}
and this is the form we want!

To summarize, the pre-limit flow $J_{t}^{\left( \lambda \right) }:\mathcal{B}%
\left( \frak{h}_{S}\right) \mapsto \mathcal{B}\left( \frak{h}_{S}\otimes 
\frak{h}_{R}\right) $ given by $J_{t}^{\left( \lambda \right) }\left(
X\right) :=U_{t}^{\left( \lambda \right) \dagger }\left( X\otimes
1_{R}\right) U_{t}^{\left( \lambda \right) }$ converges in the sense of weak
matrix elements, for fixed $X\in \mathcal{B}\left( \frak{h}_{S}\right) $, to
the limit process $J_{t}\left( X\right) =J_{t}^{\dagger }\left( X\otimes
1\right) J_{t}$. We find that $\left( J_{t}\right) _{t\geq 0}$ determines a
quantum stochastic flow on $\frak{h}_{S}\otimes \Gamma \left( L^{2}\left( 
\mathbb{R}^{+},\frak{k}\right) \right) $ and from the quantum stochastic
calculus we obtain the quantum Langevin, or stochastic Heisenberg, equation 
\begin{equation*}
dJ_{t}\left( X\right) =J_{t}\left( \mathcal{L}_{\alpha \beta }\left(
X\right) \right) \otimes dA_{t}^{\alpha \beta }.
\end{equation*}

The super-operators $\mathcal{L}_{\alpha \beta }$ are the required
Evans-Hudson maps \cite{EVHUD} $\mathcal{L}_{\alpha \beta }\left( X\right)
=XG_{\alpha \beta }+G_{\beta \alpha }^{\dag }X+G_{1\alpha }^{\dag
}XG_{1\beta }$ and these can be written in the standard form

\begin{eqnarray*}
\mathcal{L}_{11}\left( X\right) &=&\frac{1}{\gamma }\left( W^{\dagger
}XW-X\right) ; \\
\mathcal{L}_{10}\left( X\right) &=&W^{\dagger }\left[ X,L\right] ;\quad 
\mathcal{L}_{01}\left( X\right) =-\left[ X,L^{\dagger }\right] W; \\
\mathcal{L}_{00}\left( X\right) &=&\frac{\gamma }{2}\left[ L^{\dagger },X%
\right] L+\frac{\gamma }{2}L^{\dagger }\left[ X,L\right] -i\left[ X,H\right]
.
\end{eqnarray*}
In particular, $\mathcal{L}_{00}$ is a generator of Lindblad type \cite
{Lindblad}. We shall give a more detailed treatment of the convergence in
the next section.

\section{The Convergence of the Heisenberg Evolution}

We now wish to determine the limit $\lambda \rightarrow 0$ of $\left(
9.5\right) $. We have an integration over a double simplex region and the
main features emerge from examining the vacuum expectation of the product of
creation and annihilation operators. Evidently, the vacuum expectation can
be decomposed as a sum over products of two point functions and it is here
that lemma 6.1 becomes important. What must happen for a term to survive the
limit? If we have any contractions between vertices labelled by the $t$'s
then the term will vanish if the times are not consecutive. The same is true
for contractions between vertices labelled by the $s$'s. From our estimate
in the previous section, we can ignore the terms that do not comply with
this.

As a result, contractions between the $s$'s, say, will come in
time-consecutive blocks: for instance, we will typically have $m$ blocks of
sizes $r_{1},r_{2},\cdots ,r_{m}$ (these are integers 1,2,3,..., and $%
\sum_{j=1}^{m}r_{j}=n$). With a similar situation for the $t$'s, we obtain
the expansion 
\begin{eqnarray*}
&&\left\langle \phi _{1}\otimes W_{\lambda }\left( 1\right) \Phi
|\;J_{t}^{\left( \lambda \right) }\left( X\right) \;\phi _{2}\otimes
W_{\lambda }\left( 2\right) \Phi \right\rangle \\
&=&\sum_{n,\hat{n}}\left( -i\right) ^{n-\hat{n}}\sum_{m,\hat{m}%
}\sum_{r_{1},\cdots ,r_{m}}^{\sum r=n}\sum_{l_{1},\cdots ,l_{\hat{m}}}^{\sum
l=\hat{n}}\int_{\Delta _{n}\left( t\right) }ds_{n}\cdots
ds_{1}\,\int_{\Delta _{\hat{n}}\left( t\right) }dt_{\hat{n}}\cdots dt_{1}\,
\end{eqnarray*}

\begin{eqnarray*}
&&\times \left\langle \phi _{1}\right| \;\tilde{E}_{\alpha _{1}\beta
_{1}}^{\left( r_{1}\right) }\left( s_{1}^{\left( 1\right) },\dots
s_{r_{1}}^{\left( 1\right) };\lambda \right) \dots \tilde{E}_{\alpha
_{m}\beta _{m}}^{\left( r_{m}\right) }\left( s_{1}^{\left( m\right) },\dots
s_{r_{m}}^{\left( m\right) };\lambda \right) \\
&&\times X\,\tilde{E}_{\mu _{\hat{m}}\nu _{\hat{m}}}^{\left( l_{\hat{m}%
}\right) }\left( t_{1}^{\left( \hat{m}\right) },\dots ,t_{l_{\hat{m}%
}}^{\left( \hat{m}\right) };\lambda \right) \dots \tilde{E}_{\mu _{1}\nu
_{1}}^{\left( l_{1}\right) }\left( t_{1}^{\left( 1\right) },\dots
,t_{l_{1}}^{\left( 1\right) };,\lambda \right) \;\left. \phi
_{2}\right\rangle
\end{eqnarray*}
\begin{equation*}
\times \prod_{j=1}^{m}\prod_{k=1}^{r_{j}}G_{\lambda }^{\ast }\left(
s_{k+1}^{\left( j\right) }-s_{k}^{\left( j\right) }\right) \times \prod_{%
\hat{\jmath}=1}^{\hat{m}}\prod_{\hat{k}=1}^{l_{\hat{\jmath}}}G_{\lambda
}^{\ast }\left( t_{\hat{k}+1}^{\left( \hat{\jmath}\right) }-t_{\hat{k}%
}^{\left( \hat{\jmath}\right) }\right)
\end{equation*}

\begin{eqnarray}
&&\times \left\langle \Phi \right| \,\left[ a_{s_{1}^{\left( 1\right)
}}^{+}\left( \lambda \right) \right] ^{\alpha _{1}}\left[ a_{s_{r_{1}}^{%
\left( 1\right) }}^{-}\left( \lambda \right) \right] ^{\beta _{1}}\cdots %
\left[ a_{s_{1}^{\left( m\right) }}^{+}\left( \lambda \right) \right]
^{\alpha _{m}}\left[ a_{s_{r_{m}}^{\left( m\right) }}^{-}\left( \lambda
\right) \right] ^{\beta _{m}}  \notag \\
&&\left[ a_{t_{l_{\hat{m}}}^{\left( \hat{m}\right) }}^{+}\left( \lambda
\right) \right] ^{\mu _{\hat{m}}}\left[ a_{t_{1}^{\left( \hat{m}\right)
}}^{-}\left( \lambda \right) \right] ^{\nu _{\hat{m}}}\dots \left[
a_{t_{l_{1}}^{\left( 1\right) }}^{+}\left( \lambda \right) \right] ^{\mu
_{1}}\left[ a_{t_{1}^{\left( 1\right) }}^{-}\left( \lambda \right) \right]
^{\nu _{1}}\,\left. \Phi \right\rangle  \notag \\
&&\text{+ negligible terms}  \TCItag{10.1}
\end{eqnarray}
where we relabel the times as 
\begin{eqnarray*}
s_{k}^{\left( j\right) } &:&=s_{r_{1}+\cdots +r_{j-1}+k},\quad 1\leq k\leq
r_{j}; \\
t_{k}^{\left( j\right) } &:&=t_{l_{1}+\cdots +l_{j-1}+k},\quad 1\leq k\leq
l_{j};
\end{eqnarray*}
and introduce the block product of system operators 
\begin{equation*}
\tilde{E}_{\alpha \beta }^{\left( r_{j}\right) }\left( s_{1}^{\left(
j\right) },\dots s_{r_{j}}^{\left( j\right) };\lambda \right) :=\tilde{E}%
_{\alpha _{1}\beta _{1}}\left( s_{1}^{\left( 1\right) };\lambda \right) 
\tilde{E}_{11}\left( s_{2}^{\left( j\right) };\lambda \right) \cdots \tilde{E%
}_{11}\left( s_{r_{j}-1}^{\left( j\right) };\lambda \right) \tilde{E}%
_{1\beta }\left( s_{r_{j}}^{\left( j\right) };\lambda \right) .
\end{equation*}

We now examine the limit of (10.1). The estimate on the series expansion of
the Heisenberg evolute given in the previous section shows that we can
ignore the so-called negligible terms in (10.1). The limit is rather
difficult to see at this stage. However, what we can do is to recast the
expression that we claim will be the limit, 
\begin{equation}
\left\langle \phi _{1}\otimes W\left( f_{1}\otimes 1_{\left[ S_{1},T_{1}%
\right] }\right) \Psi |\,J_{t}\left( X\right) \,\phi _{2}\otimes W\left(
f_{2}\otimes 1_{\left[ S_{2},T_{2}\right] }\right) \Psi \right\rangle , 
\tag{10.2}
\end{equation}
with $J_{t}\left( X\right) =U_{t}^{\dagger }\left( X\otimes 1\right) U_{t}$,
in a more explicit form.

Recall the chaotic expansion of the process $U_{t}$ given in $\left(
8.12\right) $, the expression $(10.2)$ then becomes 
\begin{gather*}
\sum_{m,\hat{m}}\int_{\Delta _{m}\left( t\right) }\int_{\Delta _{\hat{m}%
}\left( t\right) }\sum_{r_{1},\cdots ,r_{m}}\sum_{l_{1},\cdots ,l_{\hat{m}%
}}\left( i\right) ^{\sum r-\sum l}\left( \kappa _{-}\right) ^{\sum
r-m}\left( \kappa _{+}\right) ^{\sum l-\hat{m}} \\
\times \left\langle \phi _{1}\right| \;\tilde{E}_{\alpha _{1}\beta
_{1}}^{\left( r_{1}\right) }\dots \tilde{E}_{\alpha _{m}\beta _{m}}^{\left(
r_{m}\right) }\,X\,\tilde{E}_{\mu _{\hat{m}}\nu _{\hat{m}}}^{\left( l_{\hat{m%
}}\right) }\dots \tilde{E}_{\mu _{1}\nu _{1}}^{\left( l_{1}\right) }\;\left.
\phi _{2}\right\rangle \\
\times \left\langle W\left( f_{1}\otimes 1_{\left[ S_{1},T_{1}\right]
}\right) \Psi |\,dA_{s_{m}}^{\alpha _{m}\beta _{m}}\cdots dA_{s_{1}}^{\alpha
_{1}\beta _{1}}dA_{t_{\hat{m}}}^{\mu _{\hat{m}}\nu _{\hat{m}}}\cdots
dA_{t_{1}}^{\mu _{1}\nu _{1}}\,W\left( f_{2}\otimes 1_{\left[ S_{2},T_{2}%
\right] }\right) \Psi \right\rangle .
\end{gather*}
Now the expectation between the states $W\left( f_{j}\otimes 1_{\left[
S_{j},T_{j}\right] }\right) \Psi $ can be converted into an expectation
between the Fock vacuum state $\Psi $ if we make the following replacements 
\begin{eqnarray}
dA^{11} &\rightarrow &dA^{11}+h_{2}^{\ast
}dA^{10}+h_{1}dA^{01}+h_{1}h_{2}^{\ast }dA^{10}  \notag \\
dA^{10} &\rightarrow &dA^{10}+h_{1}dA^{00}  \notag \\
dA^{01} &\rightarrow &dA^{01}+h_{2}^{\ast }dA^{00}  \notag \\
dA^{00} &\rightarrow &dA^{00}  \TCItag{10.3}
\end{eqnarray}
where $h_{j}\left( t\right) =1_{\left[ S_{j},T_{j}\right] }\left(
f_{j}|g\right) $ as in $\left( 8.1\right) $. This leads to the development 
\begin{eqnarray}
&&\sum_{m,\hat{m}}\int_{\Delta _{m}\left( t\right) }\int_{\Delta _{\hat{m}%
}\left( t\right) }\sum_{r_{1},\cdots ,r_{m}}\sum_{l_{1},\cdots ,l_{\hat{m}%
}}\left( i\right) ^{\sum r-\sum l}\left( \kappa ^{\ast }\right) ^{\sum
r-m}\left( \kappa \right) ^{\sum l-\hat{m}}  \notag \\
&&\times \left\langle \phi _{1}\right| \;\tilde{E}_{\alpha _{1}\beta
_{1}}^{\left( r_{1}\right) }\left( s_{1}\right) \dots \tilde{E}_{\alpha
_{m}\beta _{m}}^{\left( r_{m}\right) }\left( s_{m}\right) \,X\,\tilde{E}%
_{\mu _{\hat{m}}\nu _{\hat{m}}}^{\left( l_{\hat{m}}\right) }\left( t_{\hat{m}%
}\right) \dots \tilde{E}_{\mu _{1}\nu _{1}}^{\left( l_{1}\right) }\left(
t_{1}\right) \;\left. \phi _{2}\right\rangle  \notag \\
&&\times \left\langle \Psi |\,dA_{s_{m}}^{\alpha _{m}\beta _{m}}\cdots
dA_{s_{1}}^{\alpha _{1}\beta _{1}}dA_{t_{\hat{m}}}^{\mu _{\hat{m}}\nu _{\hat{%
m}}}\cdots dA_{t_{1}}^{\mu _{1}\nu _{1}}\,\Psi \right\rangle .  \TCItag{10.4}
\end{eqnarray}
where the operators $\tilde{E}_{\alpha b}^{\left( r\right) }\left( t\right) $
are given by 
\begin{equation*}
\tilde{E}_{\alpha \beta }^{\left( r\right) }\left( t\right) =\left\{ 
\begin{array}{cc}
\tilde{E}_{\alpha \beta }\left( t\right) , & r=1; \\ 
\tilde{E}_{\alpha 1}\left( t\right) \left( \tilde{E}_{11}\left( t\right)
\right) ^{r-2}\tilde{E}_{1\beta }\left( t\right) , & r\geq 2.
\end{array}
\right.
\end{equation*}
Again we note that the operators $\tilde{E}_{\alpha \beta }\left( t\right) $
have been introduced in $(8.2)$.

It remains to be shown that the limit of $(10.1)$ will be $(10.4)$. We
observe that

\begin{eqnarray}
&&\lim_{\lambda \rightarrow 0}\left\langle \phi _{1}\otimes W_{\lambda
}\left( 1\right) \Phi |\;J_{t}^{\left( \lambda \right) }\left( X\right)
\;\phi _{2}\otimes W_{\lambda }\left( 2\right) \Phi \right\rangle  \notag \\
&=&\sum_{n,\hat{n}}\left( -i\right) ^{n-\hat{n}}\sum_{m,\hat{m}%
}\sum_{r_{1},\cdots ,r_{m}}^{\sum r=n}\sum_{l_{1},\cdots ,l_{\hat{m}}}^{\sum
l=\hat{n}}\int_{\Delta _{m}\left( t\right) }ds_{m}\cdots
ds_{1}\,\int_{\Delta _{\hat{m}}\left( t\right) }dt_{\hat{m}}\cdots dt_{1}\, 
\notag \\
&&\times \left\langle \phi _{1}\right| \;\tilde{E}_{\alpha _{1}\beta
_{1}}^{\left( r_{1}\right) }\left( s_{1}\right) \dots \tilde{E}_{\alpha
_{m}\beta _{m}}^{\left( r_{m}\right) }\left( s_{m}\right) \,X\,\tilde{E}%
_{\mu _{\hat{m}}\nu _{\hat{m}}}^{\left( l_{\hat{m}}\right) }\left( t_{\hat{m}%
}\right) \dots \tilde{E}_{\mu _{1}\nu _{1}}^{\left( l_{1}\right) }\left(
t_{1}\right) \;\left. \phi _{2}\right\rangle  \notag \\
&&\times \left( \kappa _{-}\right) ^{\sum r-m}\times \left( \kappa
_{+}\right) ^{\sum l-\hat{m}}  \notag \\
&&\times \lim_{\lambda \rightarrow 0}\left\langle \Phi \right| \,\left[
a_{s_{1}}^{+}\left( \lambda \right) \right] ^{\alpha _{1}}\left[
a_{s_{1}}^{-}\left( \lambda \right) \right] ^{\beta _{1}}\cdots \left[
a_{s_{m}}^{+}\left( \lambda \right) \right] ^{\alpha _{m}}\left[
a_{s_{m}}^{-}\left( \lambda \right) \right] ^{\beta _{m}}  \notag \\
&&\left[ a_{t_{\hat{m}}}^{+}\left( \lambda \right) \right] ^{\mu _{\hat{m}}}%
\left[ a_{t_{\hat{m}}}^{-}\left( \lambda \right) \right] ^{\nu _{\hat{m}%
}}\dots \left[ a_{t_{1}}^{+}\left( \lambda \right) \right] ^{\mu _{1}}\left[
a_{t_{1}}^{-}\left( \lambda \right) \right] ^{\nu _{1}}\,\left. \Phi
\right\rangle .  \TCItag{10.5}
\end{eqnarray}
We now require the fact that 
\begin{gather*}
\lim_{\lambda \rightarrow 0}\int_{R}ds_{m}\cdots ds_{1}dt_{\hat{m}}\cdots
dt_{1}\,\left\langle \Phi \right| \,\left[ a_{s_{1}}^{+}\left( \lambda
\right) \right] ^{\alpha _{1}}\left[ a_{s_{1}}^{-}\left( \lambda \right) %
\right] ^{\beta _{1}}\cdots \left[ a_{s_{m}}^{+}\left( \lambda \right) %
\right] ^{\alpha _{m}}\left[ a_{s_{m}}^{-}\left( \lambda \right) \right]
^{\beta _{m}} \\
\left[ a_{t_{\hat{m}}}^{+}\left( \lambda \right) \right] ^{\mu _{\hat{m}}}%
\left[ a_{t_{\hat{m}}}^{-}\left( \lambda \right) \right] ^{\nu _{\hat{m}%
}}\dots \left[ a_{t_{1}}^{+}\left( \lambda \right) \right] ^{\mu _{1}}\left[
a_{t_{1}}^{-}\left( \lambda \right) \right] ^{\nu _{1}}\,\left. \Phi
\right\rangle \,f\left( s_{m},\cdots ,s_{1},t_{\hat{m}},\cdots ,t_{1}\right)
\\
=\int_{R}\left\langle \Psi |\,dA_{s_{m}}^{\alpha _{m}\beta _{m}}\cdots
dA_{s_{1}}^{\alpha _{1}\beta _{1}}dA_{t_{\hat{m}}}^{\mu _{\hat{m}}\nu _{\hat{%
m}}}\cdots dA_{t_{1}}^{\mu _{1}\nu _{1}}\,\Psi \right\rangle \,f\left(
s_{m},\cdots ,s_{1},t_{\hat{m}},\cdots ,t_{1}\right)
\end{gather*}
for $f$ continuous and $R$ a bounded region in $m+\hat{m}$ dimensions which
is the union of simplices of the type $\left( 5.2\right) $. This is readily
seen, of course, by expanding the $\Phi $-expectation as a sum of products
of two-point functions and reassembling the limit in terms of the $\Psi $%
-expectations of the processes $A_{t}^{\alpha \beta }$. This is evident from
theorems 4.2 and 6.1 quoted earlier and from the quantum It\={o} calculus 
\cite{HP}.

We therefore see that the limit form as given in (10.5) agrees with the
stated limit.

\bigskip

\noindent \textbf{Theorem (10.1)} \textit{Suppose that }$E_{\alpha \beta }$%
\textit{\ are bounded with }$K\left\| E_{11}\right\| <1$\textit{, as before.
Let }$\phi _{1},\phi _{2}\in \frak{h}_{S}$ \textit{and} $f_{1},f_{2}\in 
\frak{k}$. \textit{Then, for }$X\in \mathcal{B}\left( \frak{h}_{S}\right) $, 
\begin{eqnarray*}
&&\lim_{\lambda \rightarrow 0}\left\langle \phi _{1}\otimes W_{\lambda
}\left( 1\right) \Phi |\,J_{t}^{\left( \lambda \right) }\left( X\right)
\,\phi _{2}\otimes W_{\lambda }\left( 2\right) \Phi \right\rangle \\
&=&\left\langle \phi _{1}\otimes W\left( f_{1}\otimes 1_{\left[ S_{1},T_{1}%
\right] }\right) \Psi |\,J_{t}\left( X\right) \,\phi _{2}\otimes W\left(
f_{2}\otimes 1_{\left[ S_{2},T_{2}\right] }\right) \Psi \right\rangle .
\end{eqnarray*}

\bigskip

To summarize, the pre-limit flow $J_{t}^{\left( \lambda \right) }:\mathcal{B}%
\left( \frak{h}_{S}\right) \mapsto \mathcal{B}\left( \frak{h}_{S}\otimes 
\frak{h}_{R}\right) $ given by $J_{t}^{\left( \lambda \right) }\left(
X\right) :=U_{t}^{\left( \lambda \right) \dagger }\left( X\otimes
1_{R}\right) U_{t}^{\left( \lambda \right) }$ converges in the sense of weak
matrix elements, for fixed $X\in \mathcal{B}\left( \frak{h}_{S}\right) $, to
the limit process $J_{t}\left( X\right) =U_{t}^{\dagger }\left( X\otimes
1\right) U_{t}$. We find that $\left( J_{t}\right) _{t\geq 0}$ determines a
quantum stochastic flow on $\frak{h}_{S}\otimes \Gamma \left( L^{2}\left( 
\mathbb{R}^{+},\frak{k}\right) \right) $ and from the quantum stochastic
calculus \cite{HP} we obtain the quantum Langevin, or stochastic Heisenberg,
equation 
\begin{equation}
dJ_{t}\left( X\right) =J_{t}\left( \mathcal{L}_{\alpha \beta }\left(
X\right) \right) \otimes dA_{t}^{\alpha \beta }  \tag{10.6}
\end{equation}
where

\bigskip

The analogous result will hold for the co-evolution. Though, as mentioned
before, there is a more immediate proof using the original estimates.

\section{Conclusions}

We began with a discussion of time-ordered versus normal ordered
presentations of unitary operators relating to scattering dynamics. It is
suggestive to write the limit unitary $U_{t}$ as either 
\begin{eqnarray}
U_{t} &=&\mathbf{\vec{T}}\,\exp \left\{ -i\int_{0}^{t}ds\,E_{\alpha \beta
}\otimes \left[ \frak{a}_{s}^{\dag }\right] ^{\alpha }\left[ \frak{a}_{s}%
\right] ^{\beta }\right\} ,  \TCItag{11.1a} \\
\text{or }U_{t} &=&\mathbf{\vec{N}}\,\exp \left\{ \int_{0}^{t}ds\,L_{\alpha
\beta }\otimes \left[ \frak{a}_{s}^{\dag }\right] ^{\alpha }\left[ \frak{a}%
_{s}\right] ^{\beta }\right\} .  \TCItag{11.1b}
\end{eqnarray}
Here $\frak{a}_{t}^{\#}$ are just symbols and we mean nothing more than that 
$\left( 11.1b\right) $ is the solution of $\left( 8.10\right) $ while $%
\left( 11.1a\right) $ reminds us that we have the limit generated by a
perturbation $\Upsilon _{t}^{\left( \lambda \right) }=E_{\alpha \beta
}\otimes \left[ a_{t}^{\dag }\left( \lambda \right) \right] ^{\alpha }\left[
a_{t}\left( \lambda \right) \right] ^{\beta }$. (Formally, of course, we
might consider $\frak{a}_{t}^{\#}$ as a limiting form of the fields $%
a_{t}^{\#}\left( \lambda \right) $.)

Remarkably, these identifications $\left( 11.1a,b\right) $\ can be viewed as
presentations of $\left( 1.1\right) $ and $\left( 1.2\right) $ if we
supplement the operators $\frak{a}_{t}^{\pm }$ with the following white
noise CCR 
\begin{equation}
\left[ \frak{a}_{t},\frak{a}_{s}^{\dag }\right] =\kappa _{+}\frak{d}%
_{+}\left( t-s\right) +\kappa _{-}\frak{d}_{-}\left( t-s\right)  \tag{11.2}
\end{equation}
where $\frak{d}_{\pm }$ are future/past delta functions: we would have the
right hand side $\gamma \delta \left( t-s\right) $ if it was not for the
fact that we invariably meet with simplicial integrals. The stochastic
Schr\"{o}dinger equation $\left( 8.10\right) $ can be written as 
\begin{equation}
dU_{t}=\left[ \frak{a}_{t}^{\dag }\right] ^{\alpha }L_{\alpha \beta }U_{t}%
\left[ \frak{a}_{t}\right] ^{\beta }\,dt  \tag{11.3}
\end{equation}
which is in normal ordered form. If we understand that $\left[ \frak{a}%
_{t}^{\dag }\right] ^{\alpha }X_{\alpha \beta }\left( t\right) \left[ \frak{a%
}_{t}\right] ^{\beta }\,dt$ means $X_{\alpha \beta }\left( t\right) \otimes
dA_{t}^{\alpha \beta }$ then we recover the Hudson-Parthasarathy calculus.
The product of two quantum stochastic integrals will have to be put into
normal order, using $\left( 11.2\right) $, but this will be equivalent to
the usual quantum It\={o} rule with It\={o} table $\left( 8.9\right) $.

Alternatively, we could consider the equation $dU_{t}=-i\Upsilon
_{t}U_{t}\,dt$ with $\Upsilon _{t}=E_{\alpha \beta }\otimes \left[ \frak{a}%
_{t}^{\dag }\right] ^{\alpha }\left[ \frak{a}_{t}\right] ^{\beta }$: this is
what is suggested by $\left( 11.1b\right) $. (The Hamiltonian $\Upsilon _{t}$
plays an analogous role to the one encountered earlier for classical
stochastic Hamiltonian flows leading to $\left( 1.4\right) $.) However, the
expression $\Upsilon _{t}U_{t}$ contains terms like $\frak{a}_{t}U_{t}$
which are out of normal order and so cannot be directly interpreted in the
quantum It\={o} calculus. Nevertheless, the following purely formal
manipulations can be used \cite{GREP} 
\begin{eqnarray*}
\left[ \frak{a}_{t},U_{t}\right] &=&\left[ \frak{a}_{t},1-i\int_{0}^{t}%
\Upsilon _{s}U_{s}ds\right] =-i\kappa _{+}\int_{0}^{t}E_{1\beta }\left[ 
\frak{a}_{s}\right] ^{\beta }\frak{d}_{+}\left( t-s\right) U_{s}ds \\
&=&-i\kappa _{+}E_{11}\frak{a}_{t}U_{t}-i\kappa _{+}E_{10}U_{t},
\end{eqnarray*}
leading to 
\begin{equation}
\frak{a}_{t}U_{t}=\frac{1}{1+i\kappa _{+}E_{11}}\left\{ U_{t}\frak{a}%
_{t}-i\kappa _{+}E_{10}U_{t}\right\} .  \tag{11.4}
\end{equation}
(Similar manipulations have been performed separately for
emission-absorption and for scattering interactions in \cite{ALV}.)

By making the replacement $\left( 11.4\right) $, wherever it occurs, we
obtain a proper normal ordered form and this turns out to be precisely $%
\left( 11.3\right) $. In the classical problem for the limit of the flow
under the Hamiltonian $\left( 1.3\right) $, the canonical structure is never
lost - though we have to look to the Stratonovich calculus to see it. We
similarly have that the canonical structure is retained in the quantum
problem\ - and we even have a formal Hamiltonian $\Upsilon _{t}$- provided
that we look at things in the appropriate way.

\begin{acknowledgement}
The author is greatful to Ramon van Handel for many stimulating discussions
about the original paper that lead to several improvements and a revision of
the Heisenberg flow convergence proof.
\end{acknowledgement}

\end{document}